\documentclass[useAMS,usenatbib]{mn2e}
\usepackage{epsfig}
\usepackage{graphicx}
\usepackage{times}
\usepackage{natbib}

\newif\ifAMStwofonts
\AMStwofontstrue

\newcommand{\be}{\begin{equation}}
  \newcommand{\ee}{\end{equation}}
\newcommand{\ba}{\begin{eqnarray}}
  \newcommand{\ea}{\end{eqnarray}}
\newcommand{\brr}{\begin{array}}
  \newcommand{\err}{\end{array}}
\newcommand{\bc}{\begin{center}}
  \newcommand{\ec}{\end{center}}
\newcommand{\hm}{\,h^{-1}{\rm Mpc}}
\newcommand{\hk}{\,h^{-1}{\rm kpc}}
\newcommand{\msun}{\,h^{-1}{\rm M}_\odot}

\newcommand{\mincir}{\raise
  -2.truept\hbox{\rlap{\hbox{$\sim$}}\raise5.truept \hbox{$<$}\ }}
\newcommand{\magcir}{\raise
  -2.truept\hbox{\rlap{\hbox{$\sim$}}\raise5.truept \hbox{$>$}\ }}
\newcommand{\siml}{\raise
  -2.truept\hbox{\rlap{\hbox{$\sim$}}\raise5.truept \hbox{$<$}\ }}
\newcommand{\simg}{\raise
  -2.truept\hbox{\rlap{\hbox{$\sim$}}\raise5.truept \hbox{$>$}\ }}

\title[The effect of gas-dynamics on semi-analytic modeling of cluster
  galaxies] 
  {The effect of gas-dynamics on semi-analytic modeling of cluster galaxies} 
\author[A. Saro, et al.]  
       {A. Saro$^{1,2}$, G. De Lucia$^{3}$, K. Dolag$^3$,
         S. Borgani$^{1,2,4}$
       \\
       $^1$ Dipartimento di Astronomia dell'Universit\`a di Trieste, via
       Tiepolo 11, I-34131 Trieste, Italy (saro,borgani@oats.inaf.it)\\
       $^2$ INFN -- National Institute for Nuclear Physics, Trieste, Italy\\
       $^3$ Max--Planck--Institut f\"ur Astrophysik,
       Karl--Schwarzschild--Str. 1, D-85748 Garching bei M\"unchen, 
       Germany (gdelucia,kdolag@mpa-garching.mpg.de)\\
       $^4$ INAF, Osservatorio Astronomico di Trieste, via Tiepolo 11, I-34131
       Trieste, Italy\\ 
       }

\begin{document}

\date{Accepted ???. Received ???; in original form ???}

\pagerange{\pageref{firstpage}--\pageref{lastpage}} \pubyear{0000}

\maketitle

\label{firstpage}

\begin{abstract}
We study the degree to which non--radiative gas dynamics affects the
merger histories of haloes along with subsequent predictions from a
semi--analytic model (SAM) of galaxy formation. To this aim, we use a
sample of dark matter only and non--radiative SPH simulations of four
massive clusters. The presence of gas--dynamical processes
(e.g. ram-pressure from the hot intra--cluster atmosphere) makes
haloes more fragile in the runs which include gas. This results in a
25 per cent decrease in the total number of subhaloes at $z=0$. The
impact on the galaxy population predicted by SAMs is complicated by
the presence of `orphan' galaxies, i.e. galaxies whose parent
substructures are reduced below the resolution limit of the
simulation. In the model employed in our study, these galaxies survive
(unaffected by the tidal stripping process) for a residual merging
time that is computed using a variation of the Chandrasekhar
formula. Due to ram--pressure stripping, haloes in gas simulations
tend to be less massive than their counterparts in the dark matter
simulations. The resulting merging times for satellite galaxies are
then longer in these simulations. On the other hand, the presence of
gas influences the orbits of haloes making them on average more
circular and therefore reducing the estimated merging times with
respect to the dark matter only simulation.  This effect is
particularly significant for the most massive satellites and is (at
least in part) responsible for the fact that brightest cluster
galaxies in runs with gas have stellar masses which are about 25 per
cent larger than those obtained from dark matter only simulations. Our
results show that gas-dynamics has only a marginal impact on the
statistical properties of the galaxy population, but that its impact
on the orbits and merging times of haloes strongly influences the
assembly of the most massive galaxies.
\end{abstract}


%%%%%%%%%%%%%%%%%%%%%%%%%%%%%%%%%%%%%%%%%%%%%%%%%%%%%%%%%%%%%%%%%%%%%%%%%%%%%%%
\section{Introduction} 
\label{sec:intro} 

During the last decade, a number of observational tests of the
standard cosmological model have ushered in a new era of `precision
cosmology'. Precise measurements of angular structure in the Cosmic
Microwave Background (CMB), combined with other geometrical and
dynamical cosmological tests have constrained cosmological parameters
tightly \citep[][and references therein]{Komatsu2008} confirming
the hierarchical cold dark matter model (CDM) as the `standard' model
for structure formation. While the cosmological paradigm is well
established, our understanding of the physical processes regulating
the interplay between different baryonic components is still far from
complete, and galaxy formation and evolution remains one of the most
outstanding questions of modern astrophysics.

Different approaches have been developed in order to link the observed
properties of luminous galaxies to those of the dark matter haloes in
which they reside. Among these, semi--analytic models (SAMs) of galaxy
formation have developed into a flexible and widely used tool that
allows a fast exploration of the parameter space, and an efficient
investigation of the influence of different physical
assumptions. Computational costs are therefore reduced with respect to
hydrodynamical simulations, but this is done at the expense of an
explicit description of the gas dynamics \citep[for a recent review on
  SAMs, see][]{2006RPPh...69.3101B}. Although recent work has started
analysing the properties of the galaxy populations in hydrodynamical
simulations
\cite[e.g.][]{1996ApJ...472..460F,1999ApJ...521L..99P,2005ApJ...618...23N,2005ApJ...618..557N,2006MNRAS.373..397S,2006MNRAS.373.1265O},
the computational time is still prohibitive for simulations of
galaxies in large cosmological volumes.  In addition, the
uncertainties inherent in the physical processes at play obviously
place strong limits on the accuracy with which galaxies can be
simulated. As a consequence, these numerical studies also require an
adequate handling of `sub-grid' physics either because the resolution
of the simulation becomes inadequate to resolve the scale of the
physical process considered, or because we do not have a ``complete
theory'' of that particular physical process (which is almost always
true). It is therefore to be expected that SAMs will remain a valid
method to study galaxy formation for the foreseeable future.

In their first renditions, SAMs took advantage of Monte Carlo
techniques coupled to merging probabilities derived from the extended
Press-Schechter theory to construct merging history trees of dark
matter haloes \citep{1993MNRAS.264..201K,1994MNRAS.271..781C}. An
important advance of later years has been the coupling of
semi-analytic techniques with direct $N$-body simulations
\citep{Kauffmann_etal_1999,Benson_etal_2000}. Since dark matter only
simulations can handle large numbers of particles, such `hybrid'
models can access a very large dynamic range of mass and spatial
resolution offering, at the same time, the possibility to model the
spatial distribution of galaxies within dark matter haloes. It is also
interesting to note that there have been a number of recent studies
showing that the extended Press-Schechter formalism does not provide a
faithful description of the merger trees extracted directly from
N-body simulations
\citep{Benson_et_al_2005,Li_et_al_2007,Cole_et_al_2008}.  This might
have important consequences on the predicted properties of model
galaxies, although a detailed investigation of the influence of
analytical versus numerical merger trees on the predicted properties
of model galaxies has not been carried out yet.

A related question is whether the inclusion of the baryonic component
alters the halo dynamics with respect to a purely dark matter (DM)
simulation.  Processes like ram--pressure stripping and gas viscosity
are expected to produce a significant segregation between the
collisional and collisionless components
\citep{2001ApJ...561..708V}. These effects are likely more important
in environments characterised by high densities and large velocity
dispersions (like galaxy clusters), and are expected to change the
dynamics and the timing of halo mergers. As the merger history of
model galaxies in a SAM is essentially driven by the merger history of
its parent halo, any physical process that affects halo mergers will
influence model predictions in some measure. We note that recent work
has used merger trees from non--radiative hydrodynamic simulations
\citep[e.g.][]{cora08} to study the chemical enrichment of the
intra--cluster medium (ICM). This approach offers the advantage of
providing a three-dimensional picture of the ICM, while keeping the
advantage of exploring different physical choices with sensibly
reduced computational times with respect to hydrodynamical
simulations. The question of how SAM predictions are affected by using
merger trees from different types of simulations (e.g. DM and
hydrodynamical simulations) has, however, not been addressed.

The purpose of this paper is to quantify the effects of the presence of gas on
the merger histories of haloes, and on predictions from a galaxy formation
model.  To this aim, we have used a sample of DM-only and non--radiative
hydrodynamical simulations of four massive galaxy clusters (see
Sec.~\ref{sec:sims}).  The merger trees constructed from these simulations have
been used as input for a SAM (see Sec.~\ref{sec:SAM}), and results have been
used to carry out a careful comparison of the statistical properties of the
galaxy populations and of the formation history of the brightest cluster
galaxies (BCGs) from the two sets of simulations.

The use of non--radiative hydrodynamics is only a first step towards a
detailed comparison between SAMs and hydrodynamic simulations. A more
realistic comparison should include also gas cooling and processes
related to compact object physics, such as star formation, supernovae
feedback and supermassive black holes production and evolution. We
will present this analysis in a future work. We note that previous
work has already compared results of smooth particle hydrodynamics
(SPH) simulations and SAMs to calculate the evolution of cooling gas
during galaxy formation
(\citealt{Benson_et_al_2001,Yoshida_et_al_2002,Helly_et_al_2003}; see
also \citealt{Cattaneo_et_al_2007}), but a detailed comparison is
still lacking.

The plan of the paper is as follows. In Sec.~\ref{sec:sims} we describe the
cluster simulations used in this study, and describe the method used for the
construction of the galaxy merger trees. In Sec.~\ref{sec:SAM} we provide a
brief description of the SAM adopted, and in Sec.~\ref{sec:Res} we present the
results of our analysis. Finally, in Sec.~\ref{sec:Concl}, we summarise our
findings and give our conclusions.

%%%%%%%%%%%%%%%%%%%%%%%%%%%%%%%%%%%%%%%%%%%%%%%%%%%%%%%%%%%%%%%%%%%%%%%%%%%%%%%
\section{The simulations} 
\label{sec:sims}
 
In this study, we use a set of four simulations of massive isolated galaxy
clusters. Target haloes were identified in a DM only simulation that followed
the evolution of $512^3$ particles (with a particle mass of $7\times
10^{10}\,h^{-1}\,{\rm M}_{\odot}$) in a comoving box of size $479\,h^{-1}$Mpc
on a side \citep{2001MNRAS.328..669Y}. The simulation was carried out assuming
a flat $\Lambda$CDM cosmology with parameters: $\Omega_m = 0.3$, $h_{100}
=0.7$, $\sigma_8 = 0.9$ and $\Omega_b = 0.04$. The particles in the target
clusters and their immediate surroundings were traced back to their Lagrangian
regions and resimulated using the Zoomed Initial Condition (ZIC) technique by
\cite{TO97.2}, increasing the force and mass resolution in the region of
interest. For each halo, both a DM run and a non radiative gas run were carried
out. For the DM runs, the masses of the high--resolution DM particles is
$m_{\rm DM}\simeq 1.3\times 10^9 \msun$. In the GAS runs, the value of $m_{\rm
  DM}$ is suitably decreased so as to match the assumed cosmic baryon fraction.
The resulting mass of the gas particles is $m_{\rm gas}=1.7\times 10^8 \msun$.
In Table~\ref{t:clus}, we list the value of $M_{200}$\footnote{In this study,
  we define $M_{200}$ as the mass contained within the radius ($r_{200}$) which
  encompasses an average density of 200 times the critical density.},
$r_{200}$, and the total number of subhaloes within $r_{200}$.
 
The simulations were carried out using the TreePM--SPH code {\small GADGET-2}
\citep{2005MNRAS.364.1105S}. All GAS runs used in this study include only
non--radiative processes. The Plummer--equivalent softening length for the
gravitational force is set to $\epsilon = 5 \hk$ in physical units from $z=5$
to $z=0$, while at higher redshifts it is set to $\epsilon = 30 \hk$ in
comoving units.  The smallest value assumed for the smoothing length of the SPH
kernel is half the gravitational softening. Simulation data were stored in 93
outputs that are approximately logarithmically spaced in time down to $z\sim
1$, and approximately linearly spaced in time thereafter. Each simulation
output was analysed in order to construct merger trees of all identified
subhaloes using the software originally developed for the Millennium Simulation
project\footnote{http://www.mpa-garching.mpg.de/galform/virgo/millennium/}. We
refer to \citet{Springel_etal_2001} and to \citet{Springel_etal_2005} for a
detailed description of the substructure finder and of the merger tree
construction algorithm. In the following, we briefly summarise the main steps
of the procedure, and the changes we implemented to adapt the available
software to our simulations.

\begin{table} 
  \centering
  \caption{Some numerical information about the four clusters used in this
    study. Column 1: name of the run; Column 2: $M_{200}$, in units of
    $10^{14}\msun$; Column 3: $r_{200}$, in units of $h^{-1}$Mpc; Column 4:
    total number of subhaloes within $r_{200}$.}
  \begin{tabular}{llcr}
    Cluster name & $M_{200}$ & $r_{200}$ &
    $N_{200}$ \\  
    \hline 
    g1 DM     & 13.2  & 1.78 & 276 \\ 
    g1 GAS    & 12.2  & 1.74 & 228\\ 
    g51 DM    & 10.8  & 1.67 & 229 \\
    g51 GAS   & 10.6  & 1.66 & 200 \\
    g72 DM    & 10.9  & 1.68 & 250 \\
    g72 GAS   & 10.7  & 1.66 & 238 \\
    g8 DM     & 18.6  & 2.00 & 355 \\
    g8 GAS    & 19.4  & 2.03 & 219 \\
  \end{tabular}
  \label{t:clus}
\end{table}

For each simulation snapshot, we constructed group catalogues using a
standard friends--of--friends (FOF) algorithm with a linking length of
0.16 in units of the mean inter-particle separation. Each group was
then decomposed into a set of disjoint substructures identified as
locally overdense regions in the density field of the background main
halo. The substructure identification was performed using the
algorithm {\small SUBFIND} \citep{Springel_etal_2001}. For the
Millennium Simulation, all subhaloes with at least $20$ bound
particles were considered to be genuine substructures. In our work, we
rise this limit to at least $32$ particles. We have checked that, with
this choice, `evanescent' substructures (i.e.~objects close to the
resolution limit that occasionally appear and then disappear) are
avoided. This turns out to be important particularly for our GAS
runs. We remind the reader that {\small SUBFIND} classifies all
particle inside a FOF group either as belonging to a bound
substructure or as being unbound. The self-bound part of the FOF group
itself will also appear in the substructure list and represents what
we will refer to as the `main halo'. This particular halo typically
contains 90 per cent of the mass of the FOF group
\citep{Springel_etal_2001}.

\begin{figure*}
  \centerline{ \hbox{ \psfig{file=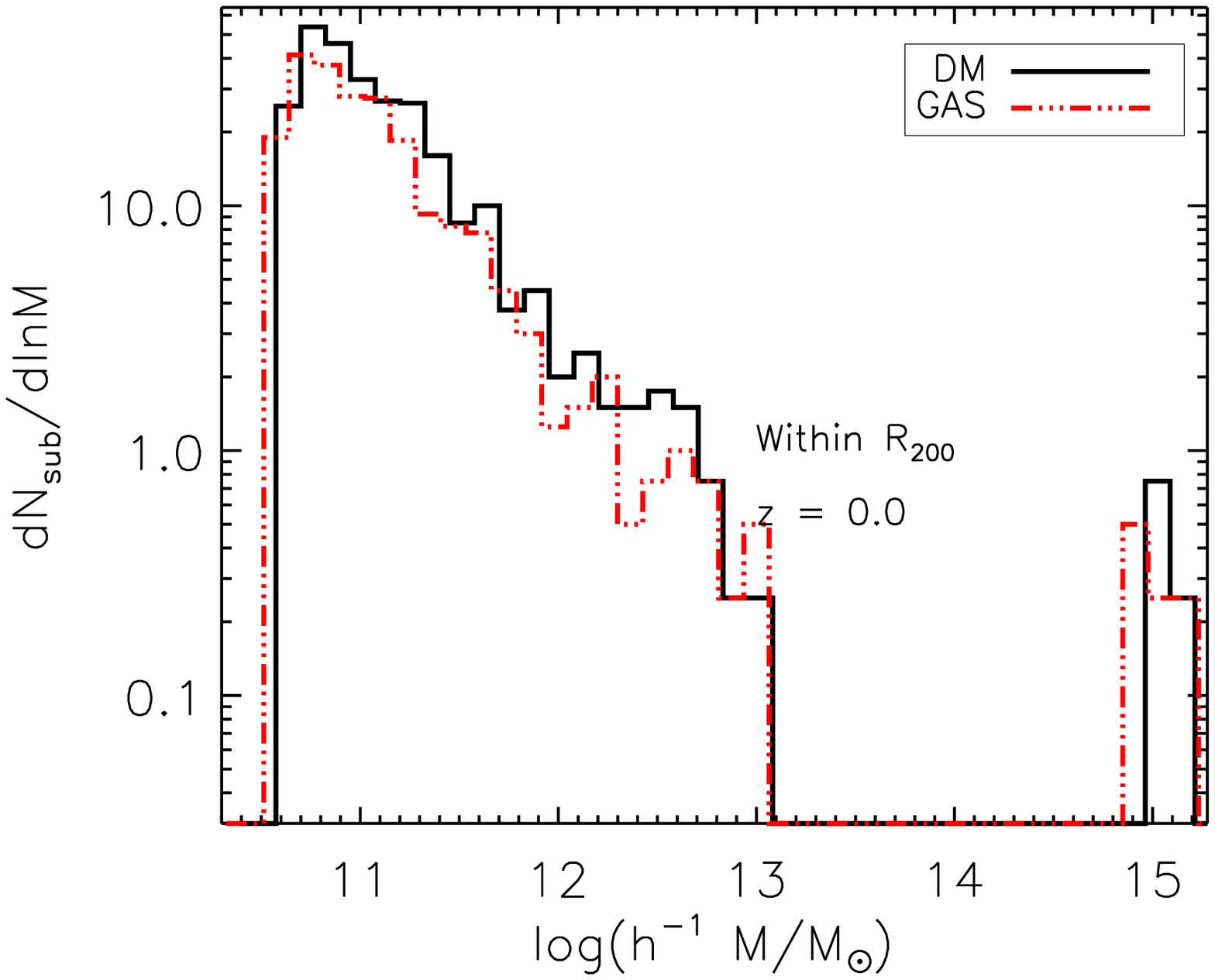,width=8.0cm}
      \psfig{file=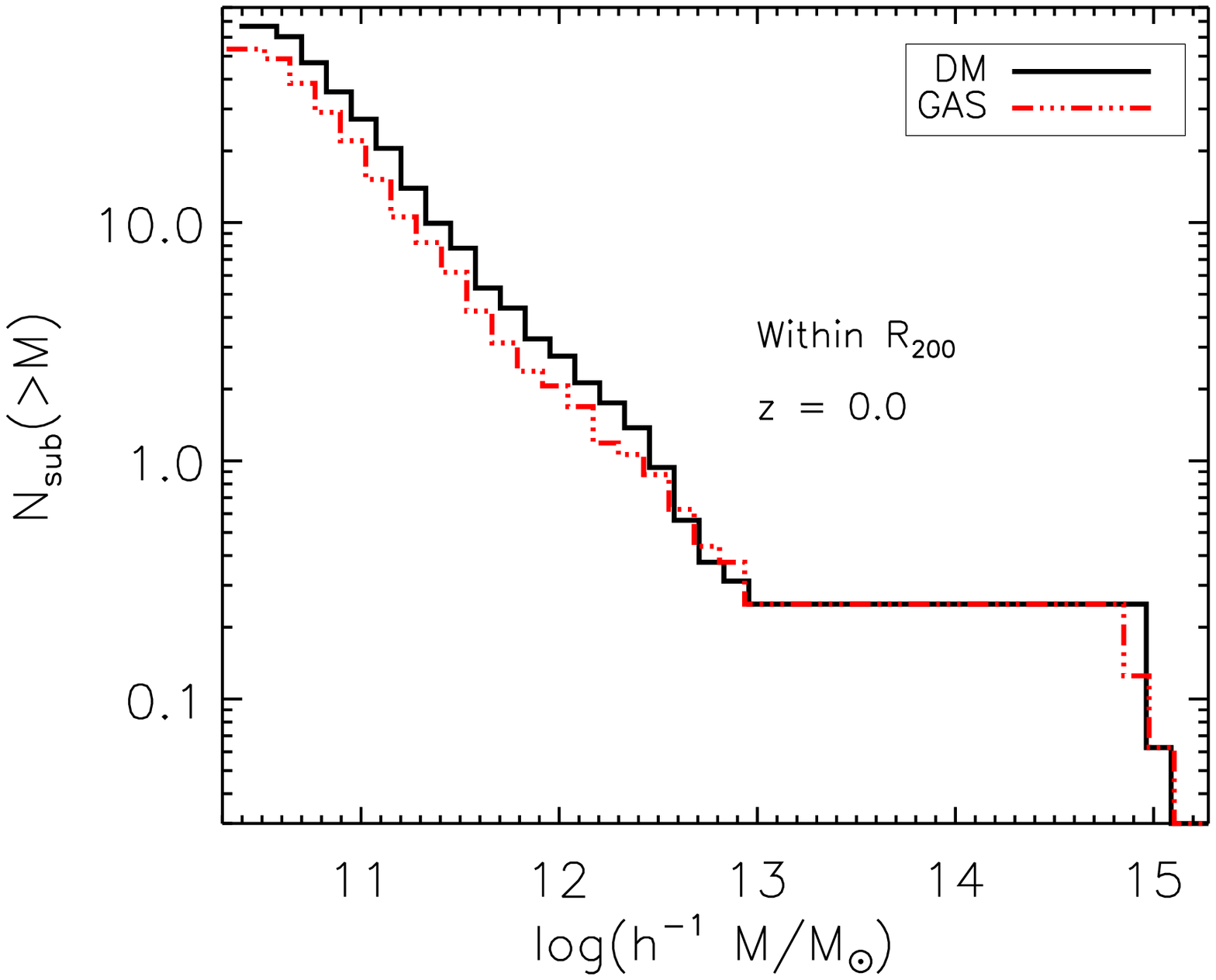,width=8.0cm} }}
  \caption{Mean differential (left) and cumulative (right) mass functions of
    all subhaloes identified within $r_{200}$ and at $z = 0$, averaged over the
    the four simulated clusters used in this study. Solid black lines are for
    the DM runs, while dot-dashed red lines are for GAS runs. For each cluster,
    we also show the main halo, using the corresponding value of $M_{200}$. For
    all other substructures, masses are given by the sum of the masses of all
    their bound particles.}
  \label{fi:MF_haloes} 
\end{figure*}

The subhalo catalogues have then been used to construct merging
histories of all self-bound structures in our simulations, using the
same procedure outlined in \citet{Springel_etal_2005}, as updated in
\citet{2007MNRAS.375....2D}.  This procedure is based on the
identification of a unique descendant for each self-bound
structure. In order to identify the descendant of a given halo, all
subhaloes in the following snapshot that contain its particles are
identified.  Particles are then counted by giving higher weight to
those that are more tightly bound in the halo under consideration. The
halo that contains the largest fraction of the most bound particles is
chosen as descendant of the halo under consideration. In our GAS runs,
the original weighting scheme used in \citet{Springel_etal_2005} leads
to a number of premature mergers for small structures. In order to
avoid this problem, we increased by a factor of one third the weight
of the most bound particles with respect to the original choice
\citep[see also][]{2008arXiv0808.3401D}. Our choice results in a
better tracing of bound structures in our GAS runs, while leaving the
results of the DM runs unaffected. The merger trees constructed as
described above represent the basic input needed for the semi-analytic
model described in Sec.~\ref{sec:SAM}.

Figure \ref{fi:MF_haloes} shows differential (left panel) and
cumulative (right panel) mass functions of the subhaloes identified at
$z = 0$ within $r_{200}$, averaged over the four simulated
clusters. We have included in these distributions the four main haloes
of the simulations, using the corresponding value of $M_{200}$ for the
mass. These correspond to the mass bins around $\sim 10^{15}\hm$ in
the differential mass function. For all other subhaloes, the mass used
in Figure \ref{fi:MF_haloes} is the sum of the masses of all their
bound particles. We will adopt this definition throughout this paper,
as well as within the semi-analytic model, whenever an estimate of the
substructure mass is needed. The left panel of Figure
\ref{fi:MF_haloes} shows that the DM mass function lies slightly but
systematically above that measured from the GAS runs. This difference
is larger than that corresponding to the shift in mass by the baryon
fraction, and it cannot be accounted for by assuming that all gas is
stripped from all subhaloes. It seems that in the non--radiative runs,
subhaloes that are stripped of their gas become both less massive more
and weakly bound \citep{2008arXiv0808.3401D}. This is probably also
the reason of the systematic difference between $M_{200}$ in DM and
GAS runs shown in Table \ref{t:clus}. The g8 cluster is an exception:
for this cluster, $M_{200}$ in the GAS run is larger than the
corresponding value from the DM run, and the number of subhaloes
within $r_{200}$ in the GAS run is much lower than the corresponding
number in the DM run. The peculiar behaviour of this cluster can be
explained by taking into account its accretion history. This is the
most massive cluster in our sample, and it did not undergo any major
merger event after $z\sim 1$. As a consequence, subhaloes in the GAS
run spent a long time in a hot, high--pressure atmosphere that can
efficiently remove their gas through ram--pressure stripping. Turning
back to Figure \ref{fi:MF_haloes}, the drop at masses $\mincir
10^{10.5} h^{-1} M_\odot$ is due to our choice of considering only
substructures with at least 32 bound particles. In the GAS runs, the
drop occurs at slightly lower masses because of the reduced value of
the gas particle mass with respect to the DM particle mass.

Although the difference is small, Figure \ref{fi:MF_haloes} shows that our GAS
runs contain less substructures than the corresponding DM runs. In the
following sections, we will analyse the impact of these differences on
prediction from a semi-analytic model of galaxy formation.

%%%%%%%%%%%%%%%%%%%%%%%%%%%%%%%%%%%%%%%%%%%%%%%%%%%%%%%%%%%%%%%%%%%%%%%%%%%%%%%
\section{The semi-analytic model}
\label{sec:SAM}

In this work, we use the semi-analytic model described in
\citet{2007MNRAS.375....2D}. We recall that the semi-analytic model we employ
builds upon the methodology originally introduced by
\citet{Kauffmann_etal_1999}, \citet{SP01.1} and \citet*{2004MNRAS.349.1101D}.
The modelling of various physical processes has been recently updated as
described in \citet{2006MNRAS.365...11C} who also included a model for the
suppression of cooling flows by `radio-mode' AGN feedback. We refer to the
original papers for details. In this study, we have assumed a Salpeter Initial
Mass Function, and a recycled gas fraction equal to 0.3.

\begin{figure*}
  \centerline{ \hbox{ \psfig{file=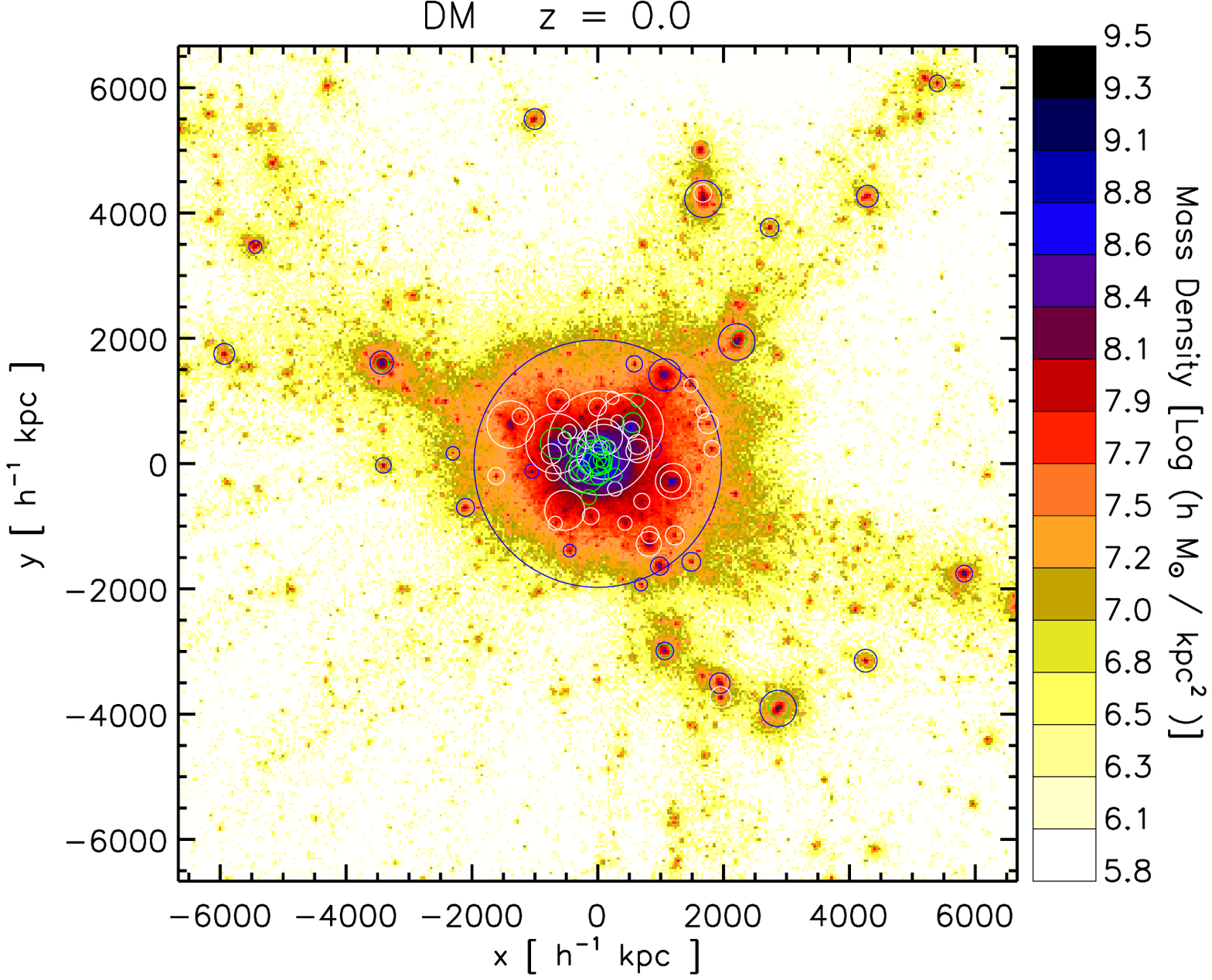,width=10.0cm}
      \psfig{file=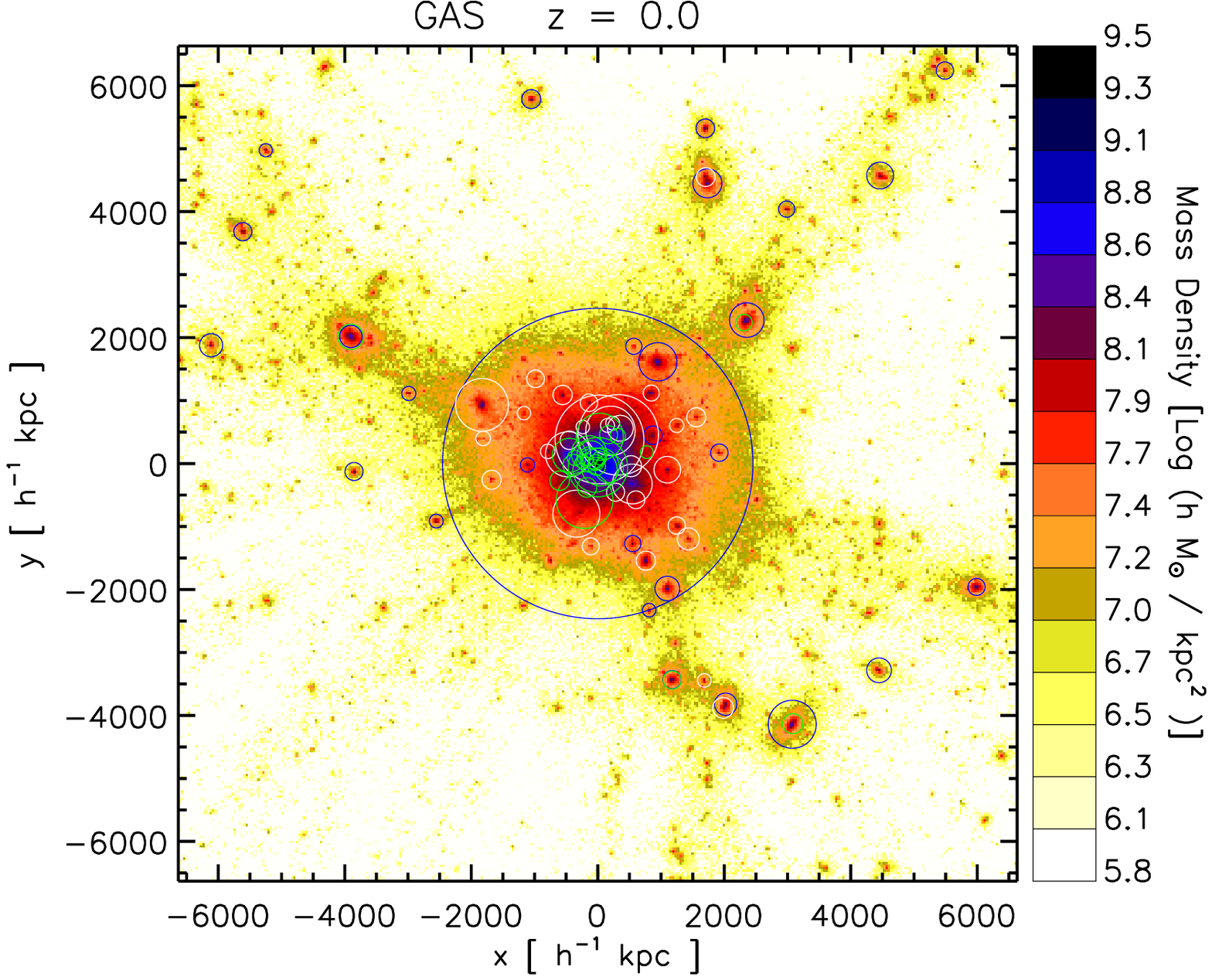,width=10.0cm} }}
  \centerline{ \hbox{ \psfig{file=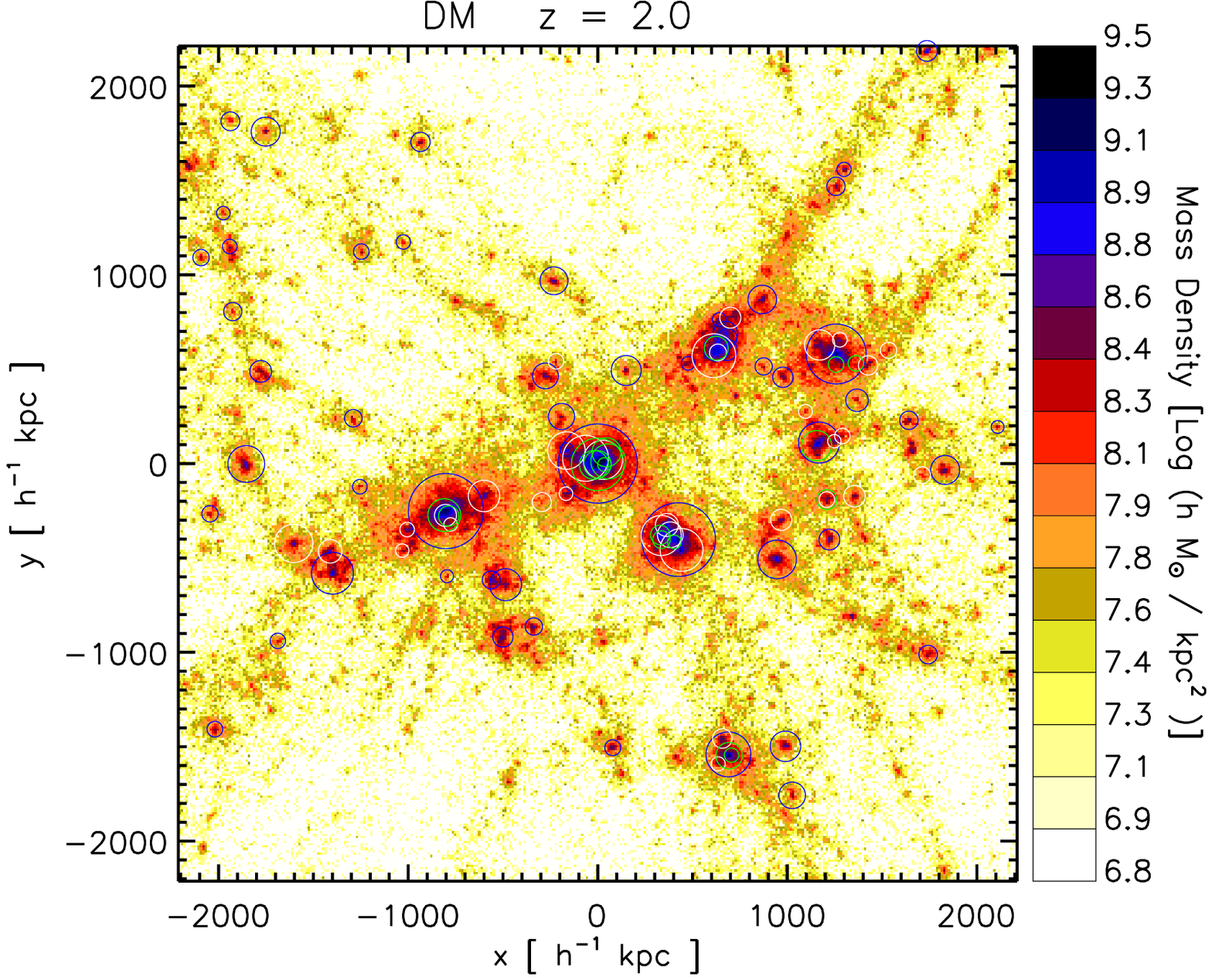,width=10.0cm}
      \psfig{file=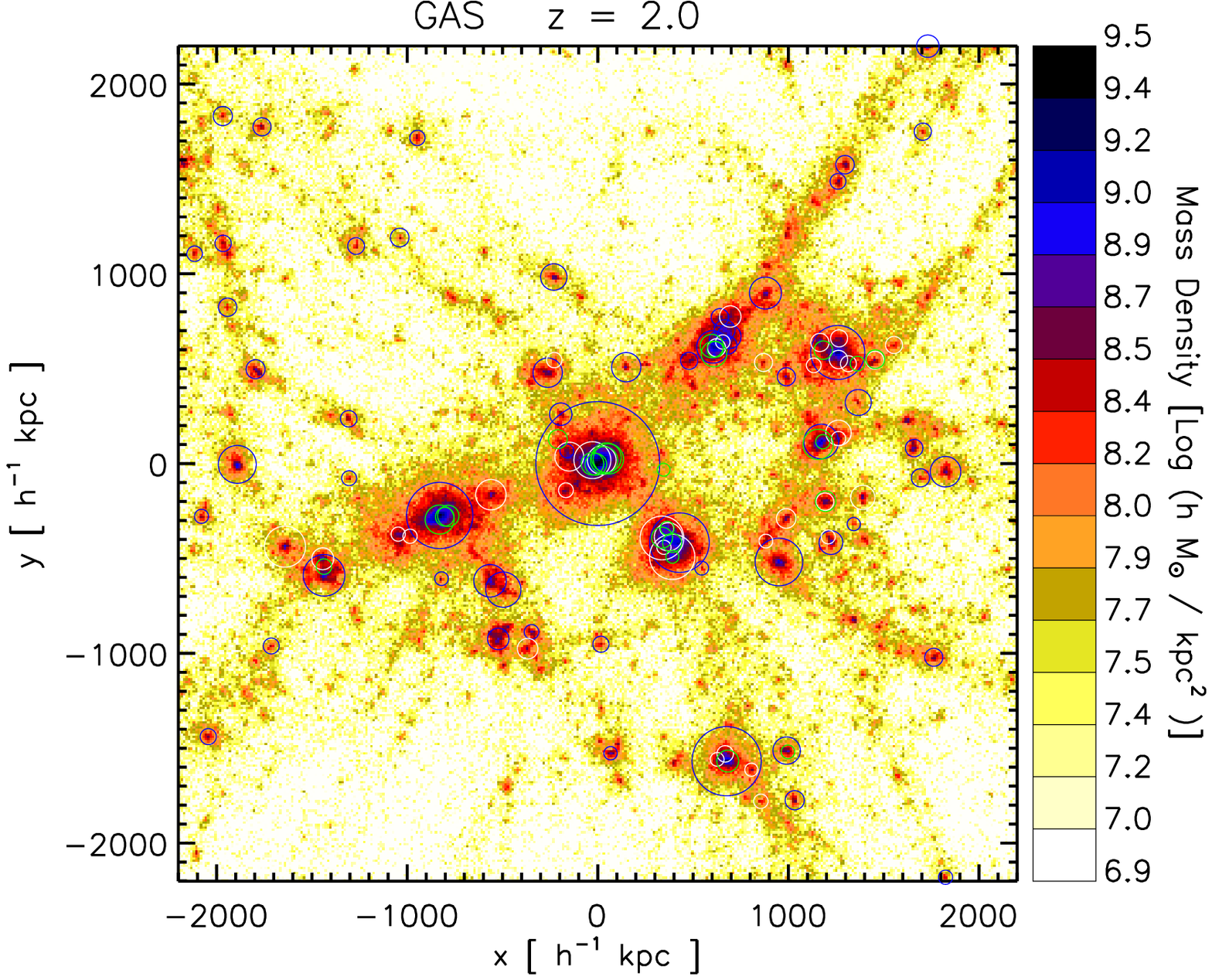,width=10.0cm} }}
  \caption{Density map of the g51 cluster for the DM run (left panels) and for
    the GAS run (right panels), at $z = 0$ (upper panels) and at $z = 2$ (lower
    panels). Positions are in comoving units. In the upper panels, the
    positions of all galaxies with stellar mass larger than $10^{11} h^{-1}
    M_\odot$ are marked by circles whose radii are proportional to the galaxy
    stellar mass. Different colours are used for different galaxy types: blue
    for Type-0, white for Type-1, and green for Type-2 galaxies. In the lower
    panels, we have marked by circles all galaxies more massive than
    $5\times10^{10} h^{-1} M_\odot$, and used the same colour-coding.}
  \label{fi:Maps} 
\end{figure*}

The semi-analytic model adopted in this study includes explicitly
dark matter substructures. This means that the haloes within which
galaxies form are still followed even when accreted onto larger
systems. As explained in \citet{SP01.1} and
\citet{2004MNRAS.349.1101D}, the adoption of this particular scheme
leads to the definition of three different `types' of galaxies. Each
FOF group hosts a `Type 0' galaxy. This galaxy is located at the
position of the most bound particle of the main halo, and it is the
only galaxy fed by radiative cooling from the surrounding hot halo
medium. All galaxies attached to dark matter substructures are
referred to as `Type 1'. These galaxies were previously central galaxy
of a halo that merged to form the larger system in which they
currently reside. The positions and velocities of these galaxies are
followed by tracing the surviving core of the parent halo. The hot
reservoir originally associated with the galaxy is assumed to be
kinematically stripped at the time of accretion and is added to the
hot component of the new main halo. Tidal truncation and stripping
rapidly reduce the mass of dark matter substructures below the
resolution limit of the simulation
\citep{2004MNRAS.348..333D,2004MNRAS.355..819G}.  When this happens,
we estimate a residual surviving time for the satellite galaxies using
the classical dynamical friction formula (see
Sec.~\ref{sec:Merg_time}), and we follow the positions and velocities
of the galaxies by tracing the most bound particles of the destroyed
substructures.  Galaxies no longer associated with distinct dark matter
substructures are referred to as `Type 2' galaxies, and their stellar
mass is assumed not to be affected by the tidal stripping that reduces
the mass of their parent haloes.

Figure \ref{fi:Maps} shows the density map of the cluster g51 from the
DM run (left panels) and from the GAS run (right panels). The
projections are colour-coded by mass density, computed within a box of
$13\,{\rm Mpc}$ comoving for the maps at $z=0$ (upper panels) and
$4.4\,{\rm Mpc}$ comoving for the maps corresponding to $z=2$ (lower
panels). The boxes corresponding to $z=0$ are centred on the most
bound particle of the main halo, while those corresponding to $z=2$
are centred on the position of the most bound particle of the main
progenitor of the cluster halo (i.e. the progenitor with the largest
mass) at the corresponding redshift. The positions of all galaxies
more massive than $10^{11} h^{-1} M_\odot$ at $z=0$ and
$5\times10^{10} h^{-1} M_\odot$ at $z=2$ are shown in projection and
marked by circles whose radii are proportional to the galaxy stellar
mass. Different colours are used for different galaxy types (blue for
Type-0, white for Type-1, and green for Type-2 galaxies). The top
panels of Figure \ref{fi:Maps} show that the massive end of the
stellar mass function at $z=0$ is dominated by Type-0 and Type-1
galaxies (blue and white circles) located within $\sim 2\,{\rm Mpc}$
from the cluster centre.  Type-2 galaxies (green circles) appear to be
more concentrated towards the centre than Type-1 galaxies (see
Sec.~\ref{sec:Dens_prof}). The brightest cluster galaxy (BCG
hereafter) in the GAS run is more massive than its counterpart in the
DM run. At $z=2$ (lower panels), the cluster is still in the process
of being assembled. In the DM run, there is no single dominant galaxy,
and the region within $\sim 2\,{\rm Mpc}$ from the main progenitor of
the BCG is characterised by the presence of other three galaxies of
similar mass. In the GAS run, the stellar mass of the main progenitor
of the BCG is already about a factor 2 larger than other massive
central galaxies in the same region. The proto-cluster regions shown
in the lower panels of Figure \ref{fi:Maps} exhibit a complex
dynamics, which witnesses the ongoing assembly of the BCG, and a
rather intense star formation activity.  This is in qualitative
agreement with observations of putative proto--cluster regions, such
as that at $z=2.16$ described by \cite{Miley_et_al_2006} and
\cite{Hatch_et_al_2008}, the so-called ``spiderweb'' galaxy. In a
forthcoming paper, we will present a detailed comparison between our
simulations and observations of proto-cluster regions.

%%%%%%%%%%%%%%%%%%%%%%%%%%%%%%%%%%%%%%%%%%%%%%%%%%%%%%%%%%%%%%%%%%%%%%%%%%%%%%%
\section{Results}
\label{sec:Res}
 In this section we will compare the mass distributions and the
 spatial distributions of the galaxies identified in the DM and GAS
 runs. We will show that, while such distributions agree quite well,
 there are differences in the final masses of the BCGs.  We argue that
 that these differences are due to the effects of gas dynamics on
 subhalo merging times and orbital distribution.

%%%%%%%%%%%%%%%%%%%%%%%%%%%%%%%%%%%%%%%%%%%%%%%%%%%%%%%%%%%%%%%%%%%%%%%%%%%%%%%
\subsection {The stellar mass function}
\label{sec:SMF}

\begin{figure*}
  \centerline{
    \hbox{
      \psfig{file=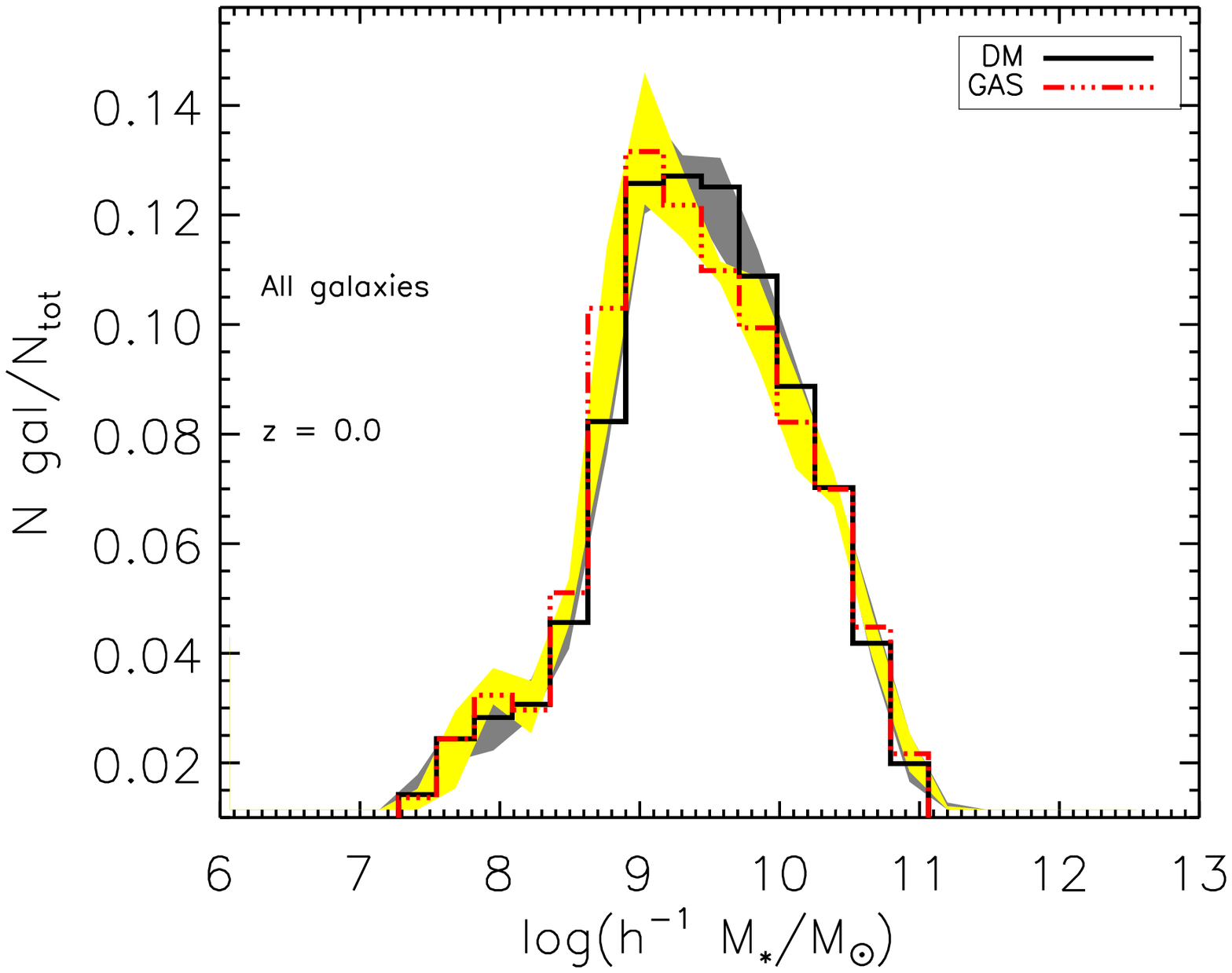,width=5.5cm} 
      \psfig{file=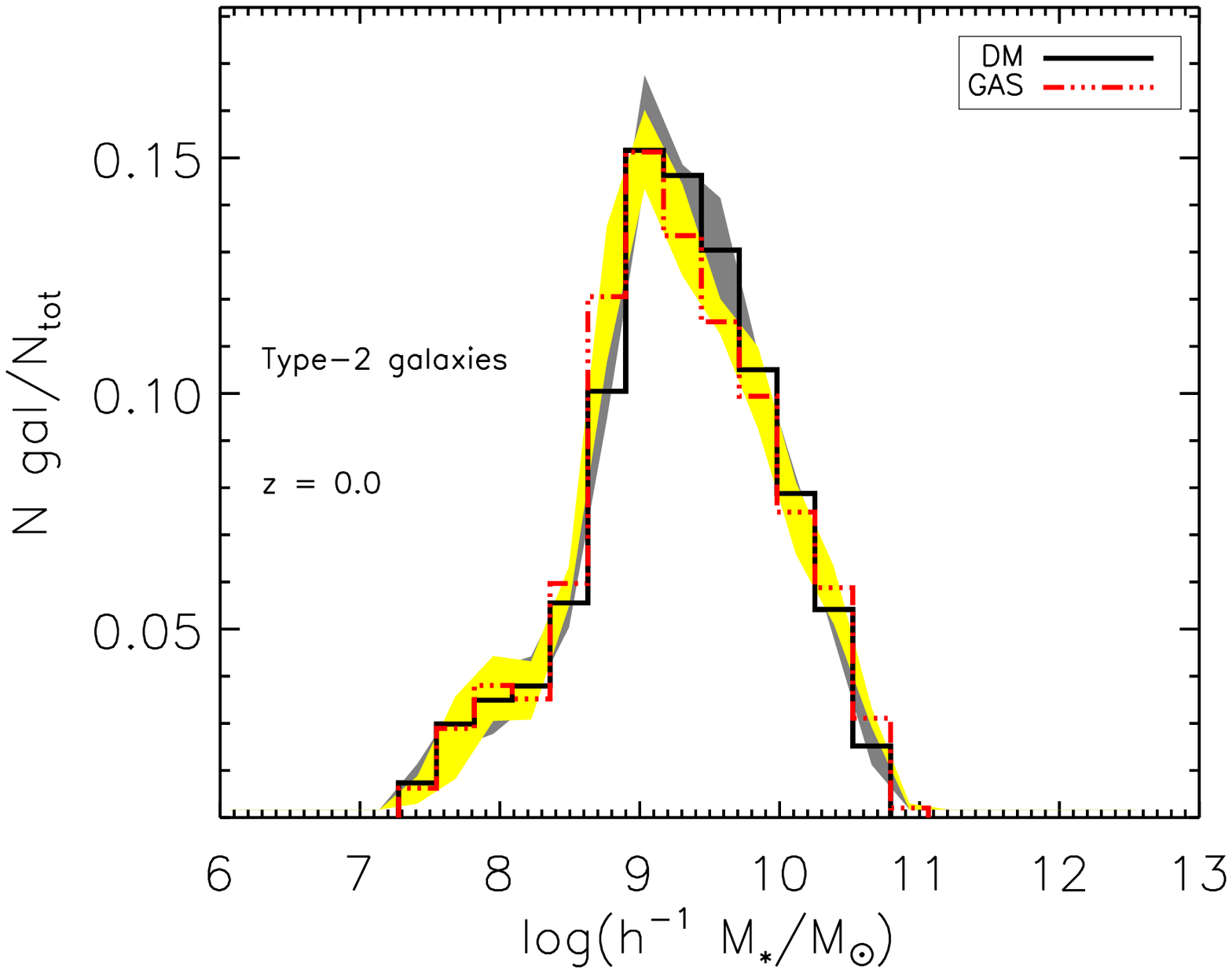,width=5.5cm} 
      \psfig{file=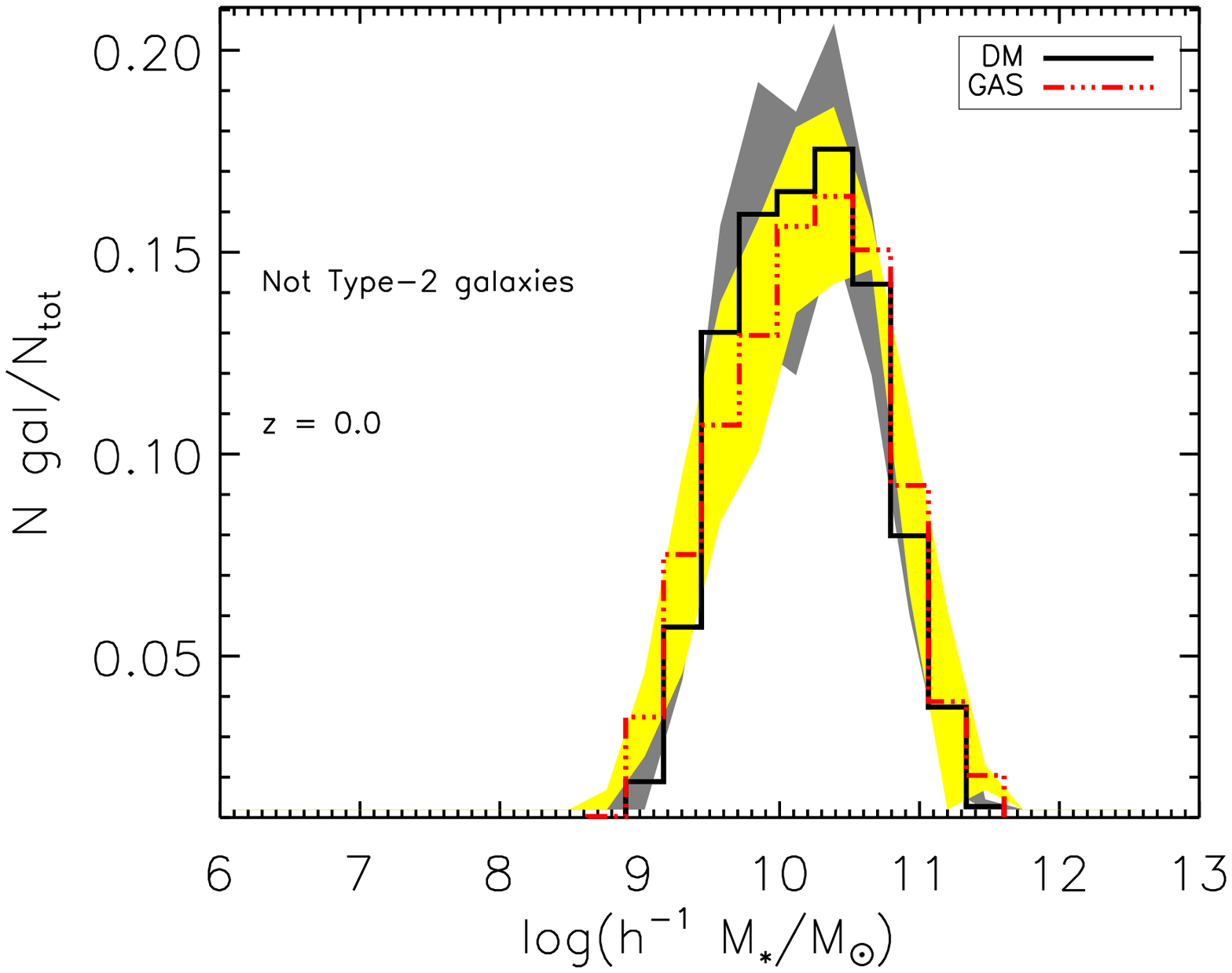,width=5.5cm} 
      }}
  \centerline{
    \hbox{
      \psfig{file=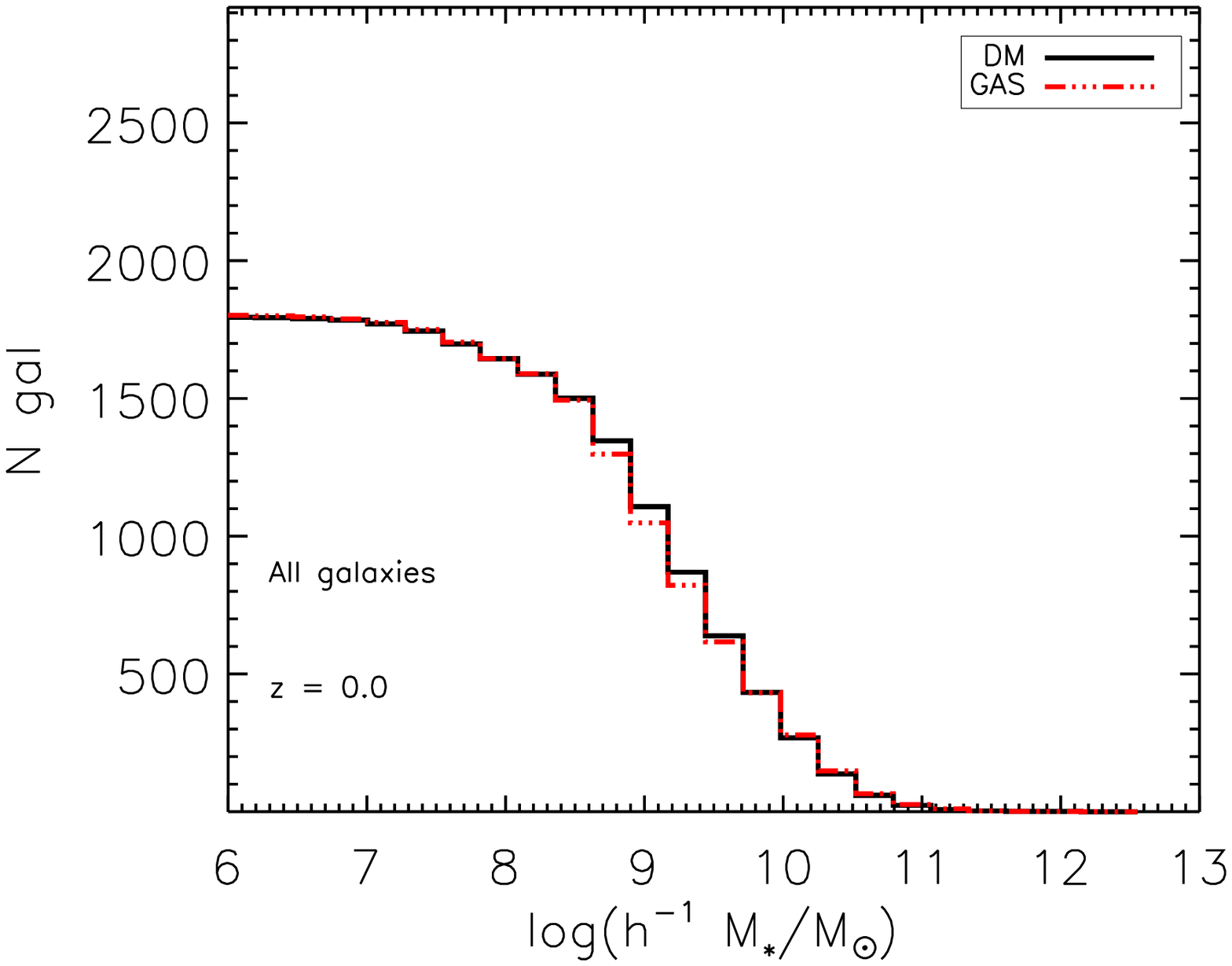,width=5.5cm} 
      \psfig{file=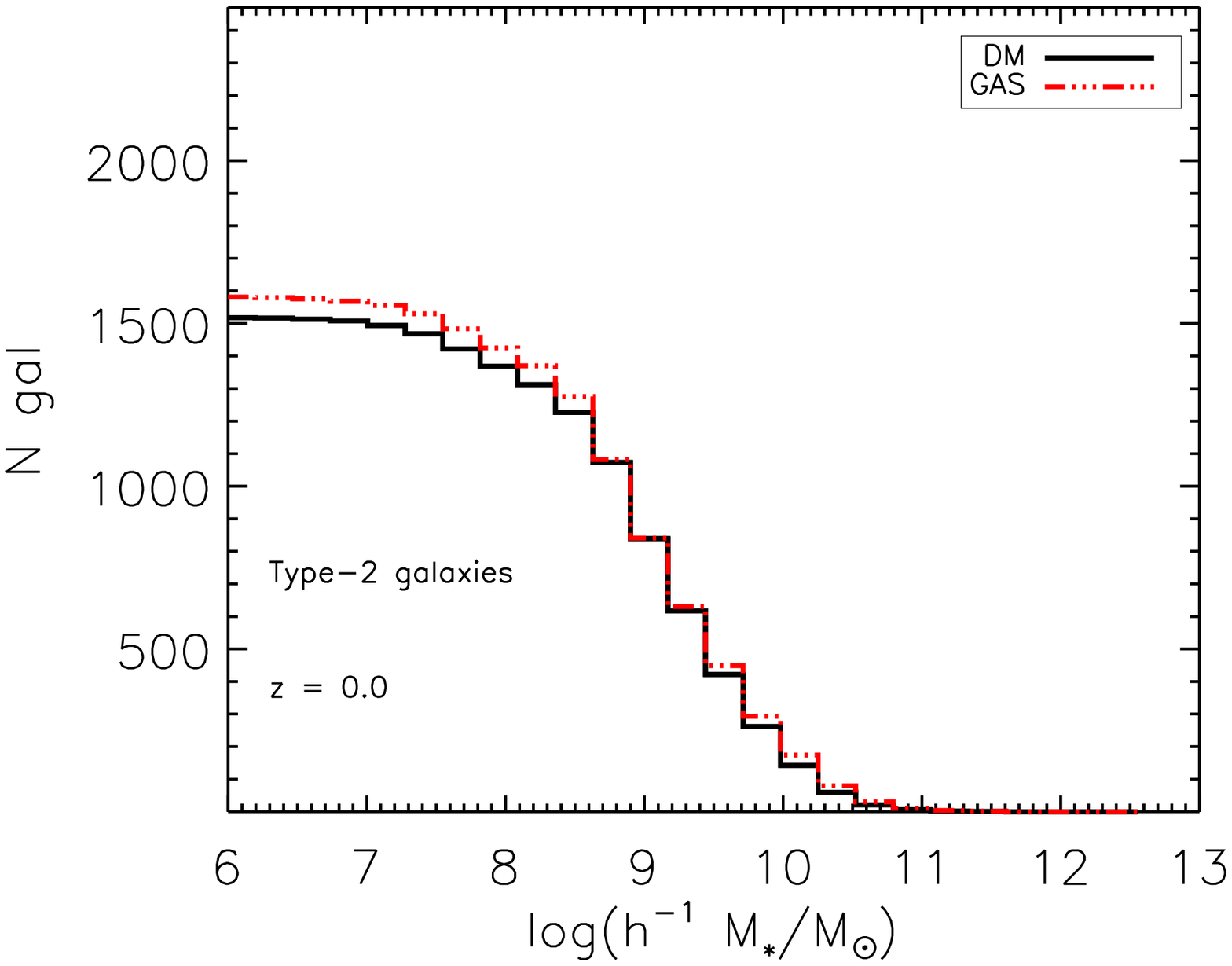,width=5.5cm} 
      \psfig{file=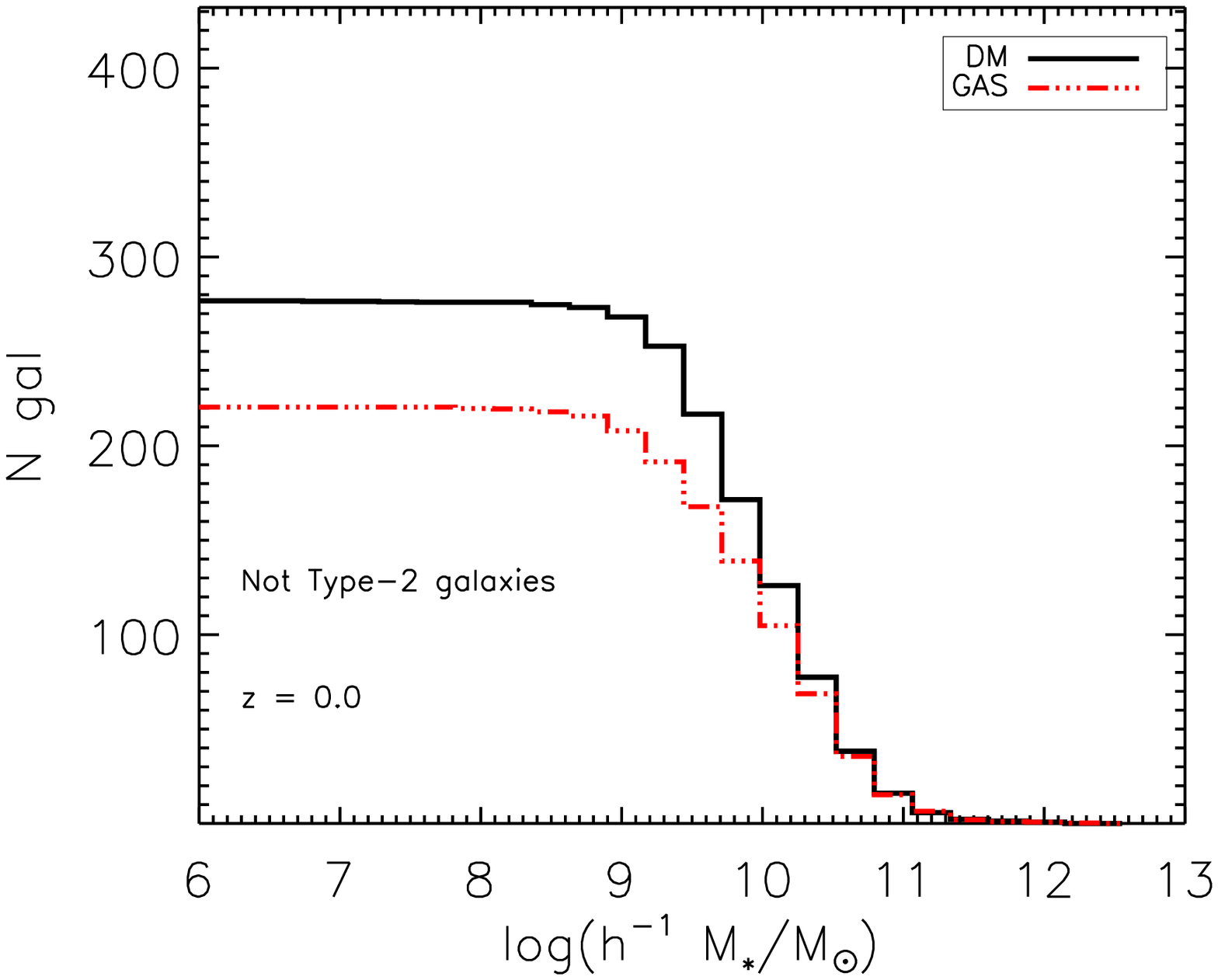,width=5.5cm} 
      }}
  \caption{Differential (top panels) and cumulative (bottom panels) stellar
    mass functions for all galaxies within $r_{200}$ at $z = 0$, in the four
    simulated clusters used in this study. The differential mass functions have
    been normalised to the total number of galaxies within $r_{200}$ in each
    cluster.  The solid histograms show the mean of the distributions from the
    four clusters, while the shaded region indicate, for each value of the
    stellar mass, the minimum and maximum number of galaxies. Solid black lines
    are for the DM runs, while dot-dashed red lines are for GAS runs. We show
    separately the stellar mass functions of the whole galaxy population (left
    panels), of the Type-2 satellite galaxy population (central panels) and the
    Type-0 and Type-1 galaxies (right panels).}
  \label{fi:MFS} 
\end{figure*}

In Sect.~\ref{sec:sims}, we have shown that the number of subhaloes in the GAS
runs is slightly lower than the corresponding number from the DM-only run (see
Figure \ref{fi:MF_haloes}).  The naive expectation is to have a total number of
galaxies in the GAS runs which is lower than the corresponding number in the DM
runs (at least Type-0 and Type-1 galaxies). Figure \ref{fi:MFS} shows that this
is indeed the case, but it also shows a number of other interesting points.

The left panels of Figure \ref{fi:MFS} show the differential (upper panel) and
cumulative (lower panel) stellar mass function for the whole galaxy population
within $r_{200}$, averaged over the four clusters used in this study. The
shaded regions indicate, for each value of the stellar mass, the minimum and
maximum number of galaxies in the simulated clusters. The agreement between the
two set of runs is quite good, but for a slight shift towards lower stellar
masses for galaxies with mass $\sim 3\times10^9\msun$. This agreement is mainly
due to the dominant contribution of the Type-2 galaxy population, whose stellar
mass function is shown in the central panels of Figure \ref{fi:MFS}. The
corresponding mass function for Type-0 and Type-1 galaxies is shown in the
right panels. In order to separate differences in the total number of galaxies
from differences in their mass distributions, the differential mass functions
shown in the upper panels of Figure \ref{fi:MFS} have been normalised to the
total number of galaxies, while the cumulative mass functions in the lower
panels indicate the un-normalised number of galaxies within $r_{200}$. We
recall that in this region, there is only one Type-0 galaxy for each cluster
(its BCG). Therefore all galaxies (but the four BCGs) shown in the right panels
are Type-1 galaxies.

The difference found in Figure \ref{fi:MF_haloes} reflects into a different
distribution and total number of Type-0 and Type-1 galaxies. This difference
is, however, compensated by the distribution and number of Type-2 galaxies,
which dominate the stellar mass function in number and represent the dominant
galaxy population at lower masses. The number of Type-2 galaxies in the GAS
runs is slightly larger, in relative terms, than in the DM runs. This small
difference is however enough to compensate the deficit of Type-1 galaxies in
the GAS runs.  As explained in Sec.~\ref{sec:sims}, we have considered as
genuine substructures all those with at least 32 bound particles.  Our
resolution limit for the galaxy stellar mass is therefore $M_{\rm star} \simeq
32 \times M_{\rm part} \times f_{\rm bar} \simeq 7 \times 10^9 h^{-1} M_\odot$
(with $f_{\rm bar} = 0.17$).  This value is close to the peak of the
differential mass functions shown in the upper panels of Figure \ref{fi:MFS}.
All galaxies below this mass limit were born in fully resolved haloes, but were
not able to transform all their baryons into stars (e.g. because their parent
halo was accreted onto a bigger system, their gas reservoir was stripped and
their star formation activity suppressed, or because they are young gas-rich
galaxies in haloes that formed relatively late).

%%%%%%%%%%%%%%%%%%%%%%%%%%%%%%%%%%%%%%%%%%%%%%%%%%%%%%%%%%%%%%%%%%%%%%%%%%%%%%%
\subsection{The number density profiles}
\label{sec:Dens_prof}

\begin{figure*}
  \centerline{ \hbox{ \psfig{file=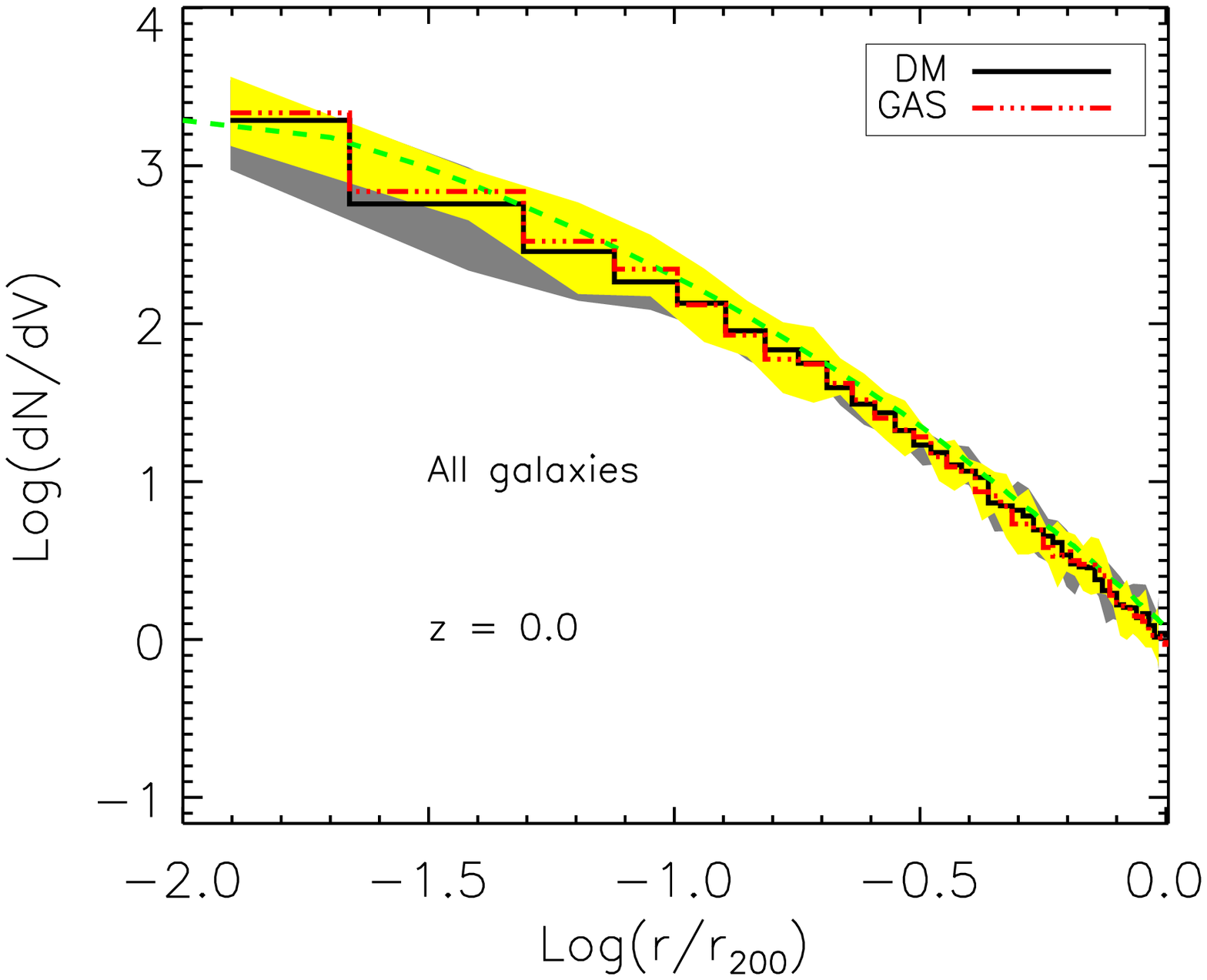,width=5.5cm}
      \psfig{file=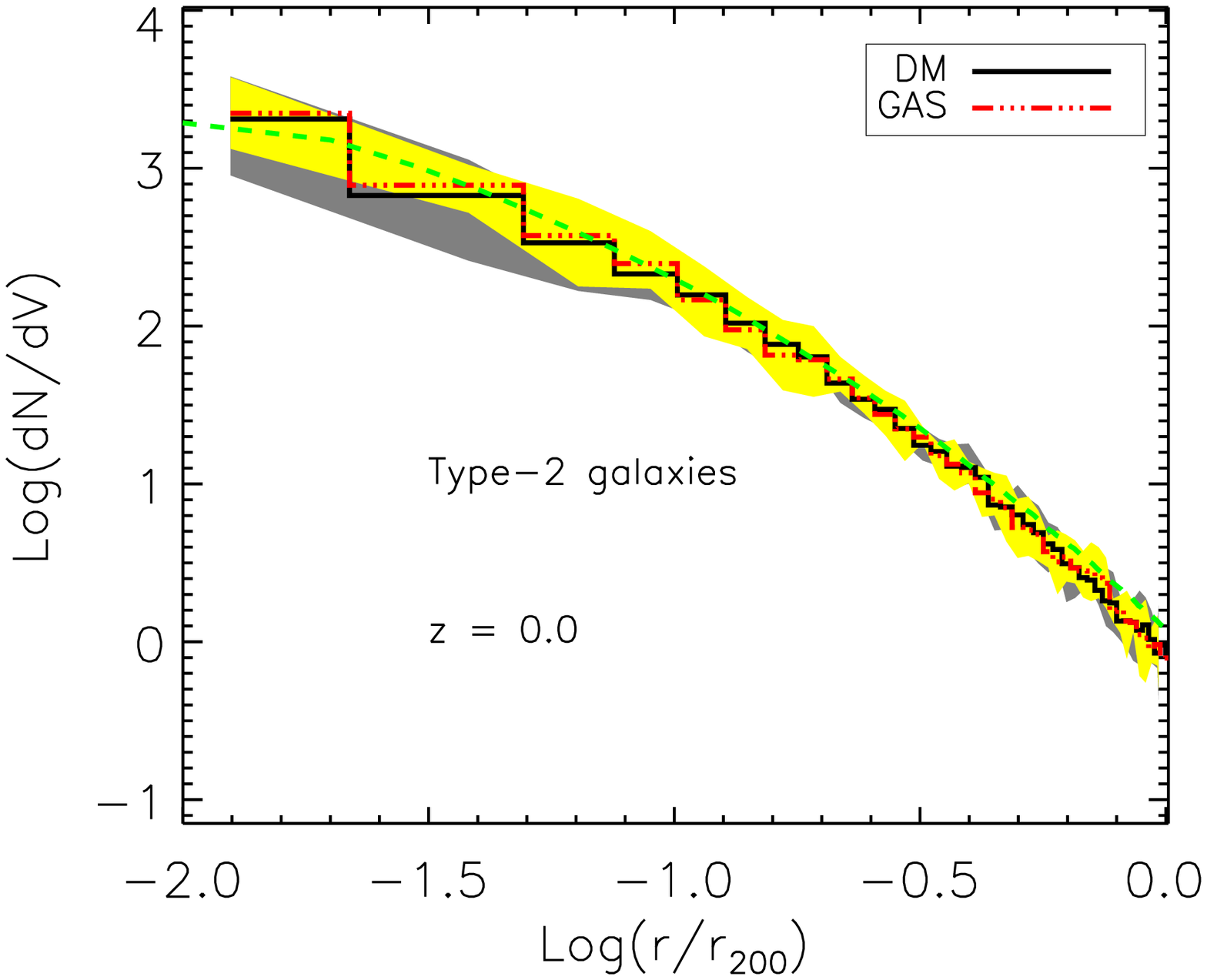,width=5.5cm}
      \psfig{file=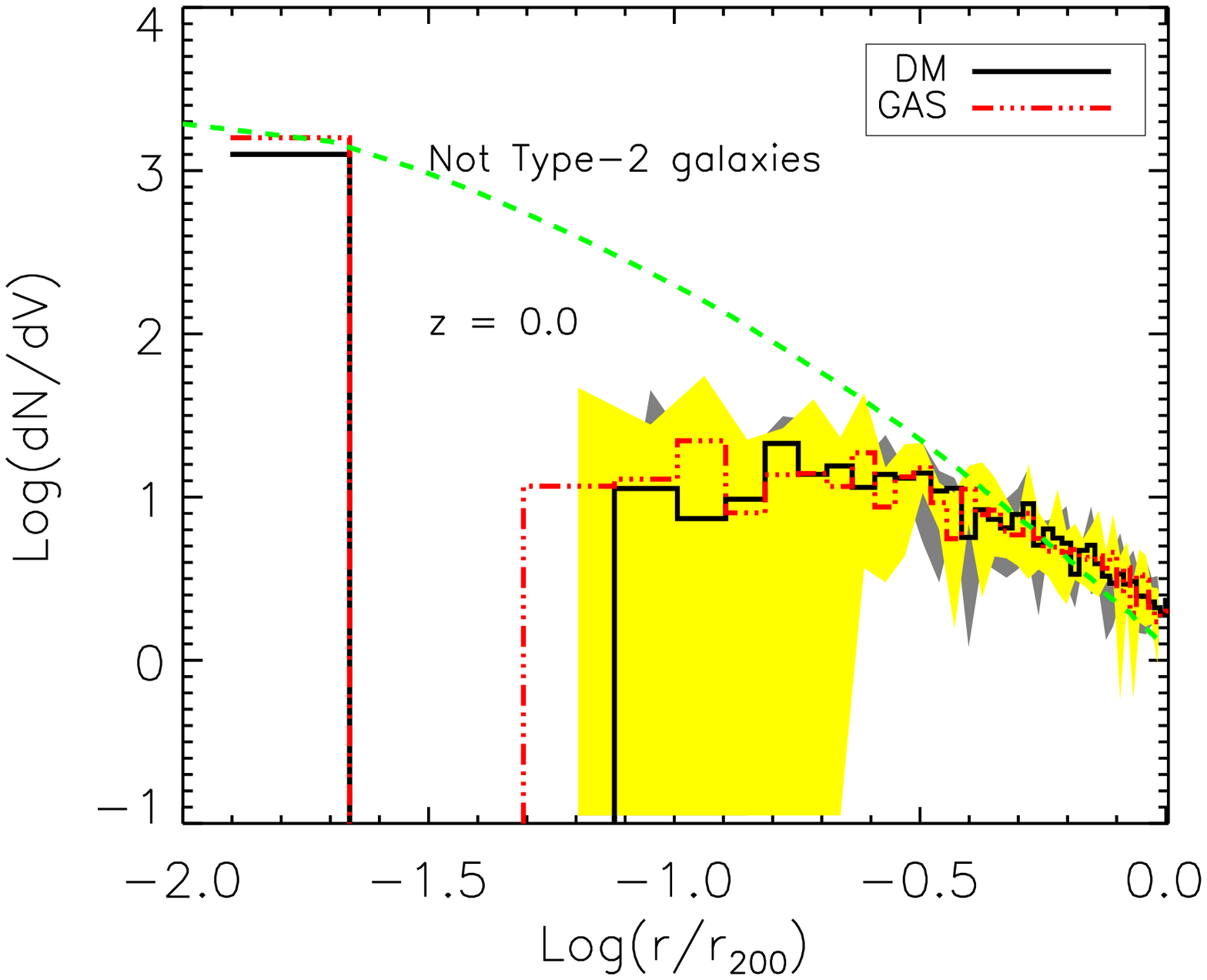,width=5.5cm} }}
  \caption{Averaged radial density of galaxies identified $z = 0$ and within
    $r_{200}$ for the four clusters used in our study. The distribution is
    normalised to the mean density of galaxies within $r_{200}$. Histograms
    show the mean of the four clusters, while their dispersion is indicated by
    the shaded regions. Solid black lines are for DM runs, while dashed-dotted
    red lines are for GAS runs. The dashed green line in each panel shows the
    mean density profile of dark matter. The three panels show separately the
    radial density of the whole galaxy population (left panel), of the Type-2
    satellite galaxy population (central panel), and of the Type-0 and Type-1
    satellite galaxies (right panel).}
  \label{fi:rad_prof}
\end{figure*}

Figure \ref{fi:rad_prof} shows the density profile of all galaxies within
$r_{200}$ from the DM (black solid lines) and the GAS (red dot-dashed lines)
runs. Solid lines show the average obtained by stacking the profiles of the
four clusters used in this study, while shaded regions show the minimum and
maximum value obtained for the simulated clusters. As for Figure \ref{fi:MFS},
we show the density profile corresponding to the whole galaxy population in the
left panel, and the contributions from Type-2 and non-Type-2 galaxies in the
central and left panels respectively.  All profiles have been normalised to the
mean galaxy density within $r_{200}$, and correspond to the galaxies identified
at $z=0$. In all panels, the dashed green line shows the average DM profile of
the simulated clusters, normalised to match the density profile of the galaxies
in the inner bin.
 
The galaxy density profile is dominated by the Type-2 galaxy population at all
radii, and follows very nicely the underlying DM profile, in agreement with
what found by \citet{Gao_etal_2004}. By definition, the central Type-0 galaxies
populate the innermost bin in Figure \ref{fi:rad_prof}. The right panel of this
figure shows that Type-1 galaxies tend to avoid the central cluster regions,
where they are efficiently destroyed by the intense tidal field of the parent
halo. The radial profile of Type-1 galaxies is `anti-biased' relative to the
dark matter profile in the inner regions, as expected from studies of dark
matter substructures \citep{Ghigna_etal_2000,2004MNRAS.349.1101D}. 

The agreement between the DM and GAS runs is quite good. The only notable
difference is a small shift towards the centre for the positions of Type-1
galaxies in the GAS runs. We have verified, however, that this difference is
due to a single galaxy which is found closer to the centre in the GAS run of
the cluster g72. The shift is due to the influence of the gas on the orbit of
substructures, as we will discuss in the following.

The good agreement for the radial distribution of Type-2 galaxies in the two
sets of runs used in this study is not obvious. We recall that the positions of
Type-2 galaxies are given by the updated positions of the particles that were
the most bound particles of the parent substructure, before their masses were
reduced below the resolution limit of the simulation. The agreement between the
DM and GAS runs therefore implies that the presence of gas in the simulation
does not significantly alter the distribution of those particles, which trace
the spatial distribution of DM particles. 

%%%%%%%%%%%%%%%%%%%%%%%%%%%%%%%%%%%%%%%%%%%%%%%%%%%%%%%%%%%%%%%%%%%%%%%%%%%%%%%
\subsection{Merging times}
\label{sec:Merg_time}

In the previous section, we have shown that the cluster galaxy population
resulting from the model employed in this study is dominated in number by
Type-2 galaxies. Model predictions for this galaxy population are very
sensitive to the residual merging times that are assigned to Type-2 galaxies
when their parent dark matter subhaloes are stripped below the resolution limit
of the simulation.

These merging times, which regulate for how long a Type-2 galaxy keeps its
identity before merging with the central galaxy of its own halo, are computed
using the following implementation of the \cite{1943ApJ....97..255C} dynamical
friction formula:
\begin{equation}
  T_{\rm merge}\,=\,1.17\times{D^2\times V_{\rm virial}\over
    \log({M_{\rm main} \over M_{\rm sat}} + 1) \times G \times
    M_{\rm sat}}\,.
  \label{eq:merg}
\end{equation}
In the above equation, $D$ is the distance between the merging halo and the
centre of the structure on which it is accreted, $V_{\rm virial}$ is the
circular velocity of the accreting halo at the virial radius, $M_{\rm sat}$ is
the mass associated with the merging satellite, and $M_{\rm main}$ is the mass of
the accreting halo. All quantities entering in Eq.~\ref{eq:merg} are computed
at the last time the merging satellite can be associated with a resolved
dark matter substructure. We note that satellite galaxies can merge either with
Type-0 or with Type-1 galaxies, but the majority of the mergers occur between 
Type-2 and Type-0 galaxies. In this case, $M_{\rm main}$ is $M_{200}$ of the
accreting halo, while in the case of a merger between a Type-2 and a Type-1
galaxy, $M_{\rm main}$ is given by the sum of the masses of all bound particles
associated with the accreting halo.

We note that Eq.~\ref{eq:merg} is adapted from the original
formulation derived by Chandrasekhar in the approximation of an
orbiting point mass satellite in a uniform background mass
distribution. It is also worth reminding the reader that our
formulation of dynamical friction does not include any dependence on
the orbital distribution. Furthermore, $M_{\rm sat}$ in our SAM
formulation does not include the mass associated with the stars and to
the cold inter-stellar medium of the galaxy in the merging
substructure. We have verified, however, that by taking into account
this baryonic component does not alter significantly our results.

Two recent papers \citep{2008MNRAS.383...93B,2008ApJ...675.1095J} have studied
merging time-scales using N-body and hydro/N-body simulations. Both studies
have pointed out that a formulation similar to that given in Eq.~\ref{eq:merg}
systematically underestimates merging time-scales, although they derived new
fitting formulae which differ in a number of details. In this work, we are only
concerned with differences due to the application of the same formula to
different runs, while we plan to come back to the validity of the Chandrasekhar
formula in a future work. We also note that, in the standard application of the
dynamical friction formula, quantities related to the merging satellite are
computed at the time at which the satellite crosses the virial radius of the
accreting halo, while in our case satellites are traced until their mass is
reduced below the resolution limit of the simulation by tidal stripping.
Merging times are computed at this time, so that $D$ is the distance of the
merging satellite (the position of its most bound particles) to the centre of
the accreting halo, and it can be larger or smaller than the virial radius of
this halo.

\begin{figure}
  \centerline{\hbox{\psfig{file=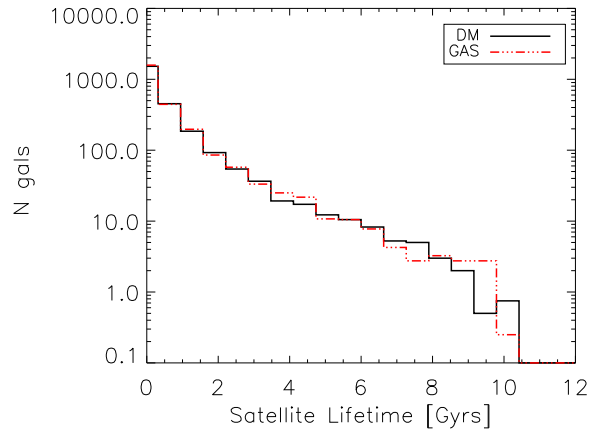, width=8.0cm}}}
  \caption{Distribution of lifetimes of Type-1 galaxies. The lifetime is
    defined here as the time elapsed between the last time the galaxy was a
    Type-0 galaxy and the time the substructure mass was reduced below the
    resolution limit of the simulation (the galaxy becomes a Type-2). Results
    from the four simulated clusters have been stacked together. The solid
    black line is for the DM runs, and the red dashed line corresponds to the
    GAS runs.}
  \label{fi:Life_t1} 
\end{figure}

In Sec.~\ref{sec:sims}, we have shown that the number of haloes in the GAS runs
is approximately equal or slightly lower than the number of haloes in the DM
runs. Given this difference in the number of haloes, an excess of Type-2
galaxies in the GAS runs can have two possible explanations: a shorter lifetime
of substructures or longer merging times assigned to Type-2 galaxies. In order
to test which of these two alternative explanations applies, we turn to our
simulation results. 

\begin{figure*}
  \centerline{ \hbox{ \psfig{file=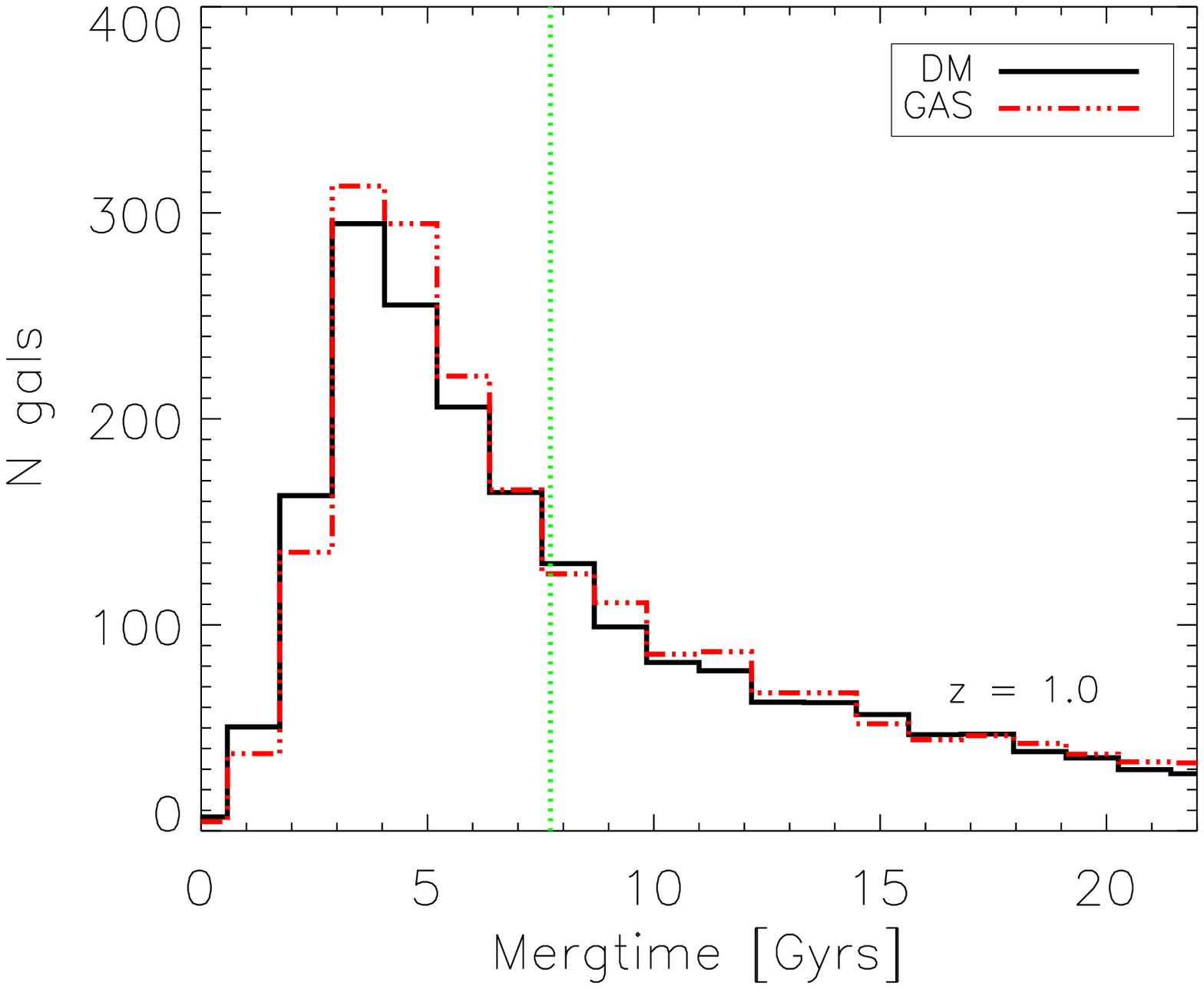,width=5.5cm}
      \psfig{file=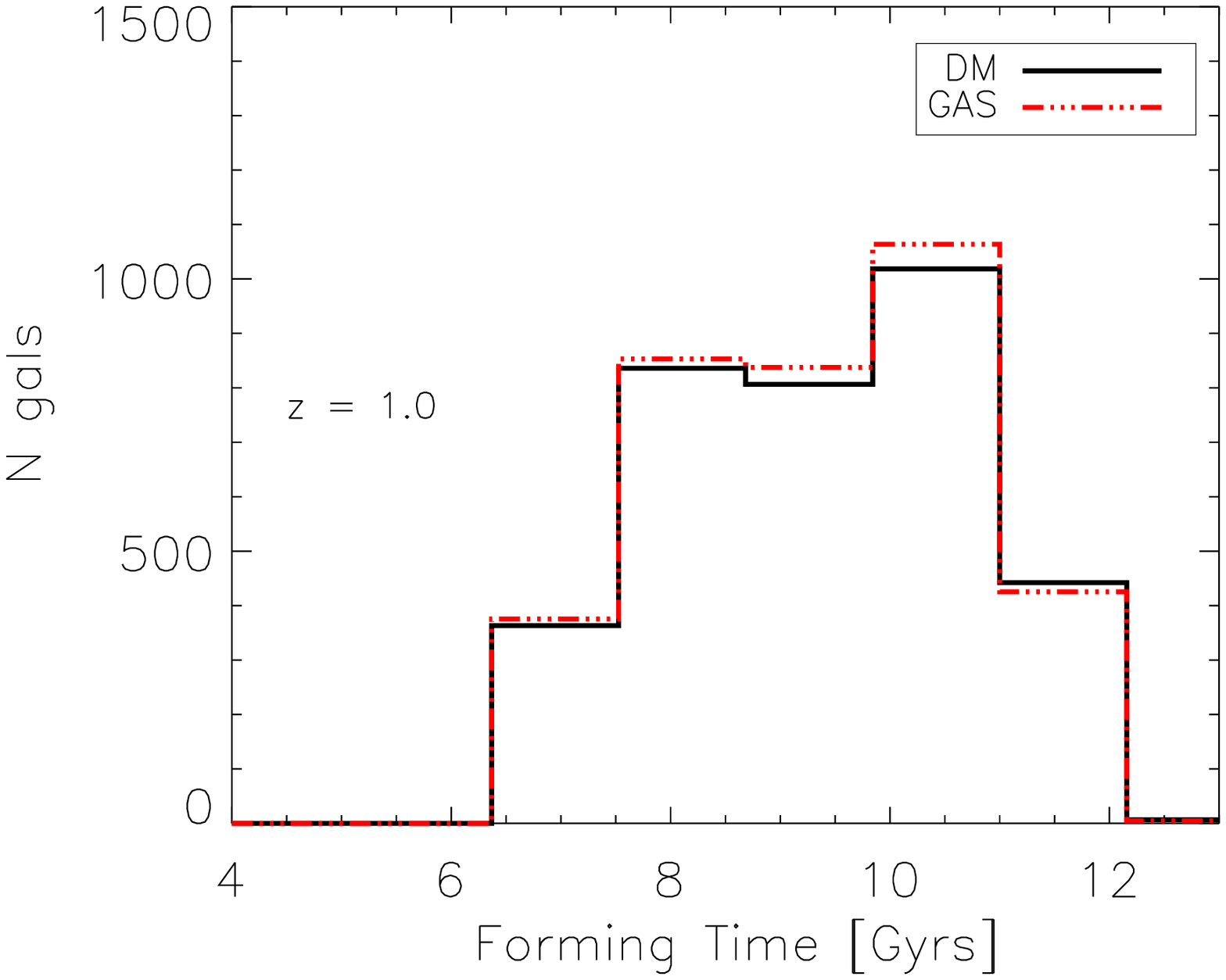,width=5.5cm}
      \psfig{file=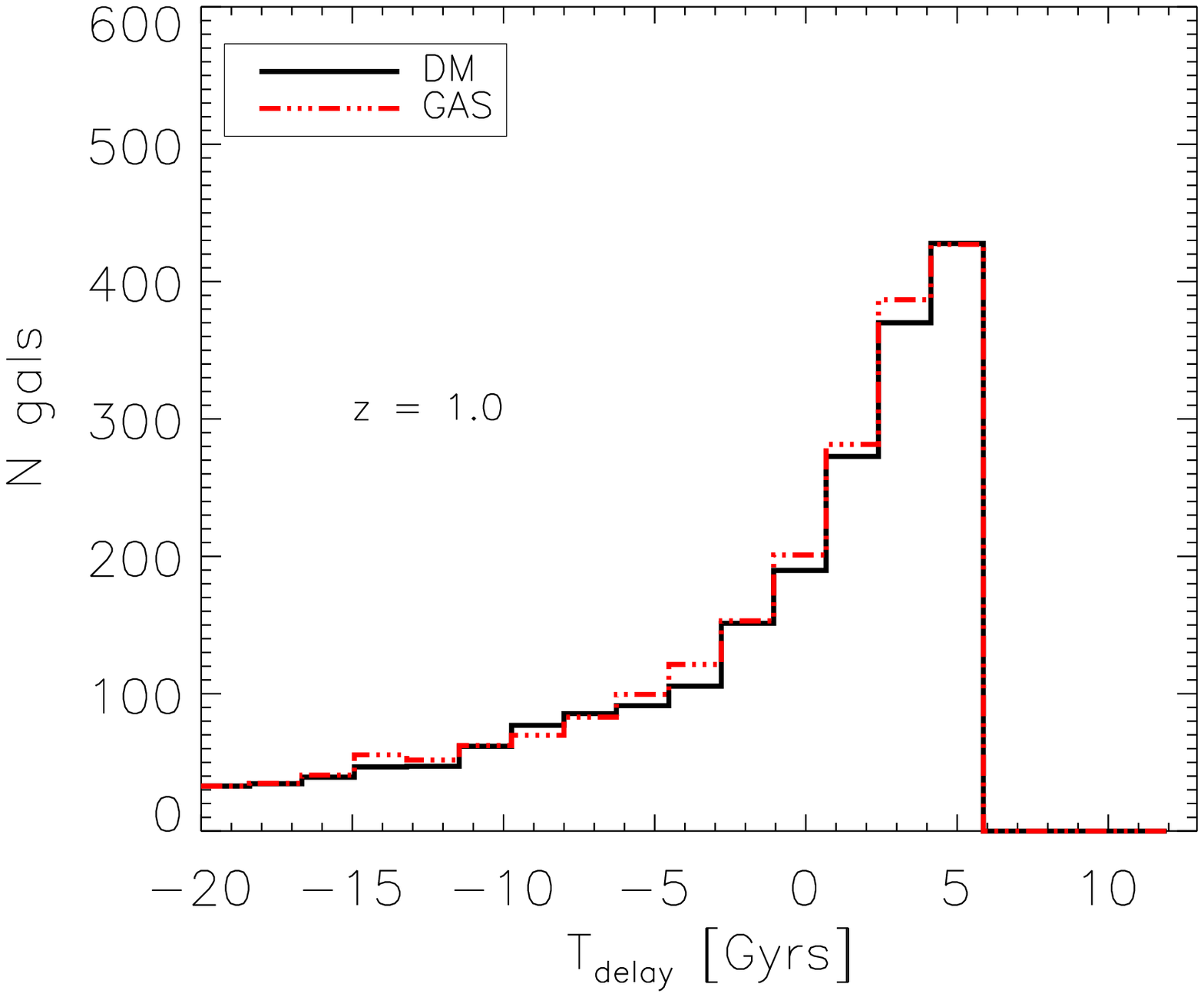,width=5.5cm} }}
  \caption{Left panel: distributions of the merging times assigned to the
    Type-2 galaxies identified at $z=1$ in the four simulated clusters. The
    solid black line shows the result from the DM runs, while the dot-dashed
    red line is for the GAS runs. The vertical green dotted line marks the
    value of the lookback time at $z = 1$. All galaxies whose merging time is
    smaller than this value will merge by $z = 0$. Central panel: distribution
    of the formation times of the Type-2 galaxies, i.e. the lookback times when
    the galaxies first become Type-2. Right panel: distribution of the
    differences between the formation and the merging times ($T_{\rm delay}$)
    of the Type-2 galaxies identified $z = 1$ (see text).}
  \label{fi:Merg_plot} 
\end{figure*}

Figure \ref{fi:Life_t1} shows the distribution of the lifetimes of subhaloes in
the DM (solid black line) and in the GAS simulations (dashed red line). For
each subhalo, its lifetime is computed as the time elapsed between the last
time it was identified as a main halo (i.e. hosting a Type-0 galaxy at its
centre) and the time it merged (i.e. its mass dropped below the resolution
limit of the simulation). For this figure and for those in the remainder of
this section, we have stacked the results from the four clusters used in this
study. Figure \ref{fi:Life_t1} shows that the distribution of lifetimes of
Type-1 galaxies does not differ significantly in the DM and in the GAS runs.
The excess of Type-2 galaxies must therefore be ascribed to different merging
times associated with them in the two runs, with merging times expected to be
systematically longer in the GAS runs. In order to verify our expectation, we
show in the left panel of Figure \ref{fi:Merg_plot} the distribution of the
merging times which were assigned to Type-2 galaxies identified at $z = 1$.
The figure shows that there is an excess of Type-2 galaxies with merging times
shorter than $\sim 3$~Gyr in the DM runs, and a corresponding excess of
galaxies with merging times larger than the same value in the GAS runs. The
central panel of Figure \ref{fi:Merg_plot} shows the formation times of the
Type-2 galaxies identified at $z = 1$, i.e. the lookback times when the
galaxies become Type-2 for the first time. The distributions are very similar,
confirming that these Type-2 galaxies form on average at the same time in the
GAS and in the DM runs. The right panel of Figure ~\ref{fi:Merg_plot} shows the
distribution of $T_{\rm delay}$, which is defined as the difference between the
formation times of the Type-2 galaxies identified at $z=1$ and the merging times
they get assigned. According to this definition, the integral of this
distribution between a given $T_{\rm delay}$ and infinity gives the total
number of Type-2 galaxies at $z = 1$ that have merged since lookback time
$T_{\rm delay}$. The fact that the distribution for the GAS runs lies above
that of the DM runs for negative values of $T_{\rm delay}$ indicates that the
excess of Type-2 galaxies in the GAS runs will continue at least in the next
five Gyrs after the present time.

Figure ~\ref{fi:Merg_plot} therefore proves that the excess of Type-2 galaxies
in the GAS runs (shown in the central panels of Figure \ref{fi:MFS}) is due to
the fact that these galaxies get assigned longer merging times in the GAS runs
with respect to the corresponding merging times assigned to Type-2 galaxies in
the DM runs. It is interesting to ask what is the origin of these differences.
This can be done by considering all quantities entering Eq.~\ref{eq:merg}.

\begin{figure*}
  \centerline{ \hbox{ \psfig{file=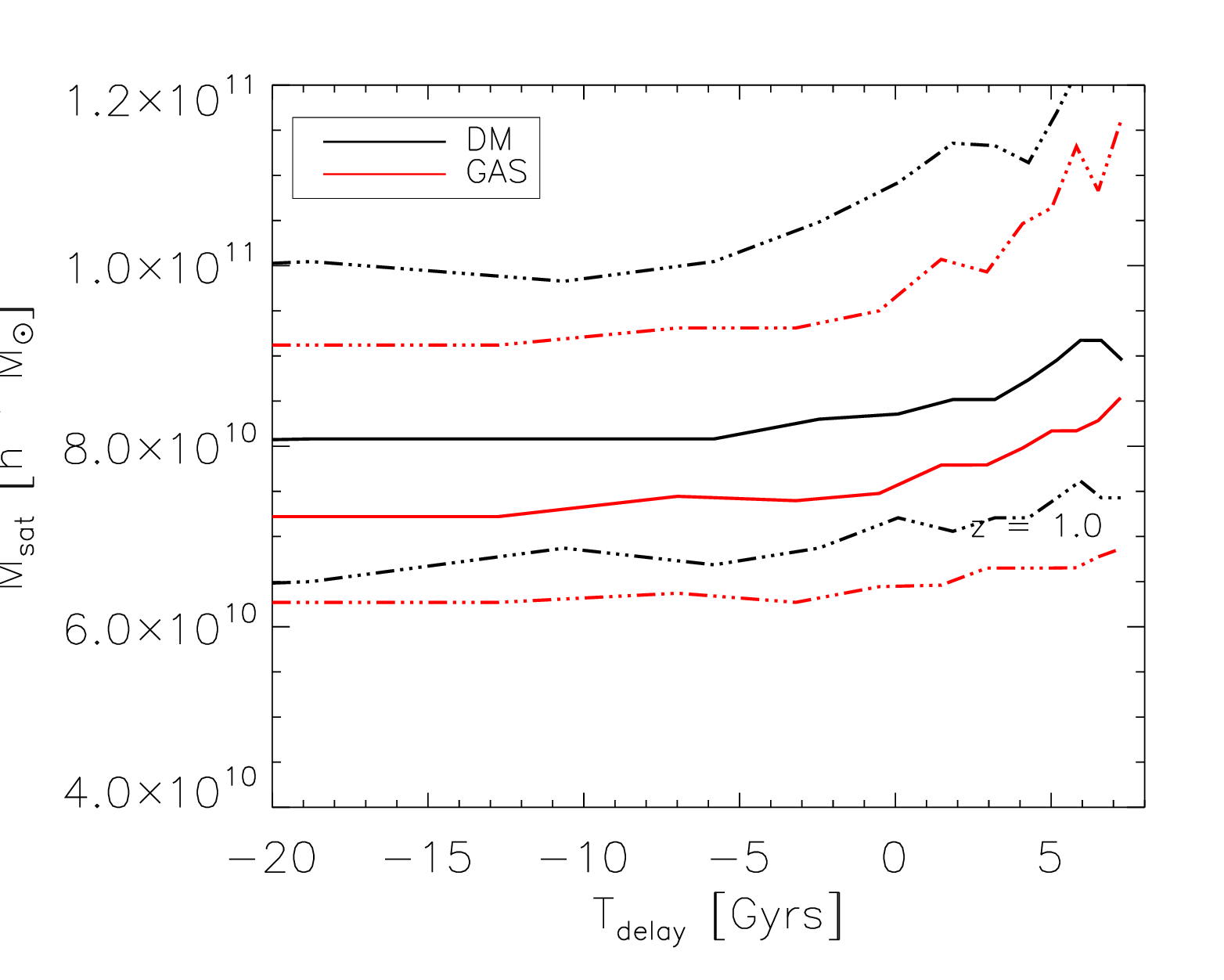,width=9.0cm}
      \psfig{file=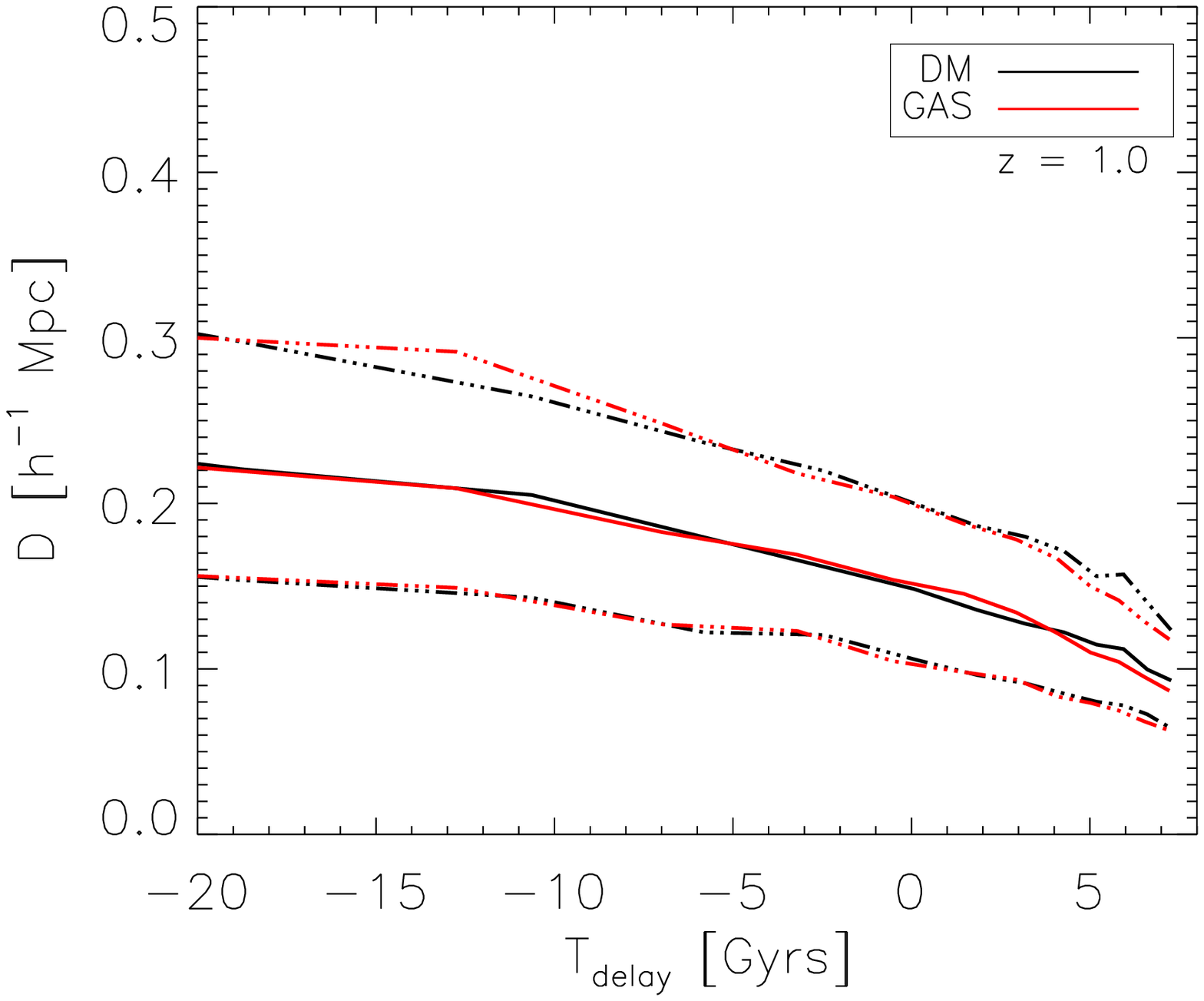,width=9.0cm} }} \centerline{ \hbox{
      \psfig{file=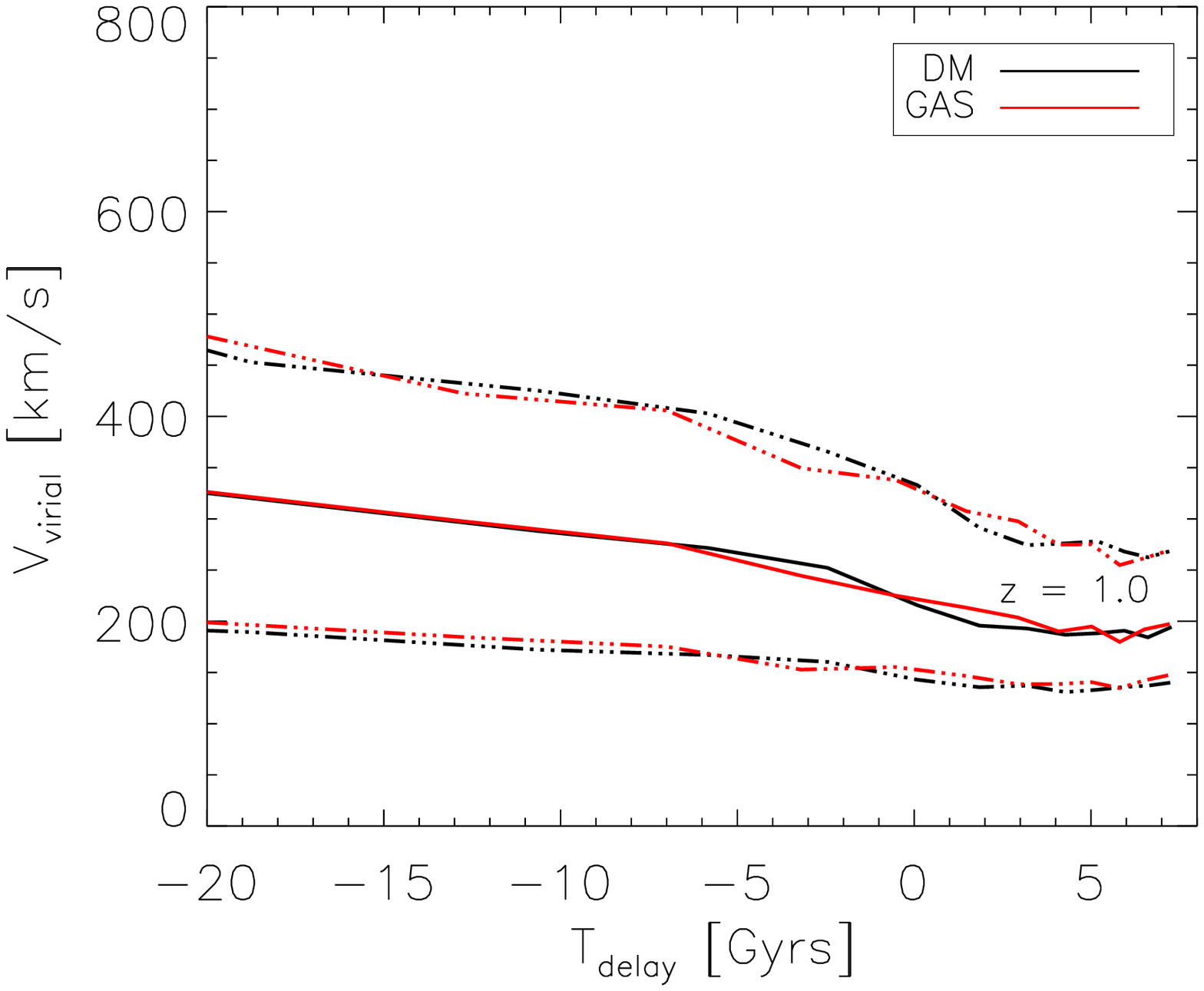,width=9.0cm}
      \psfig{file=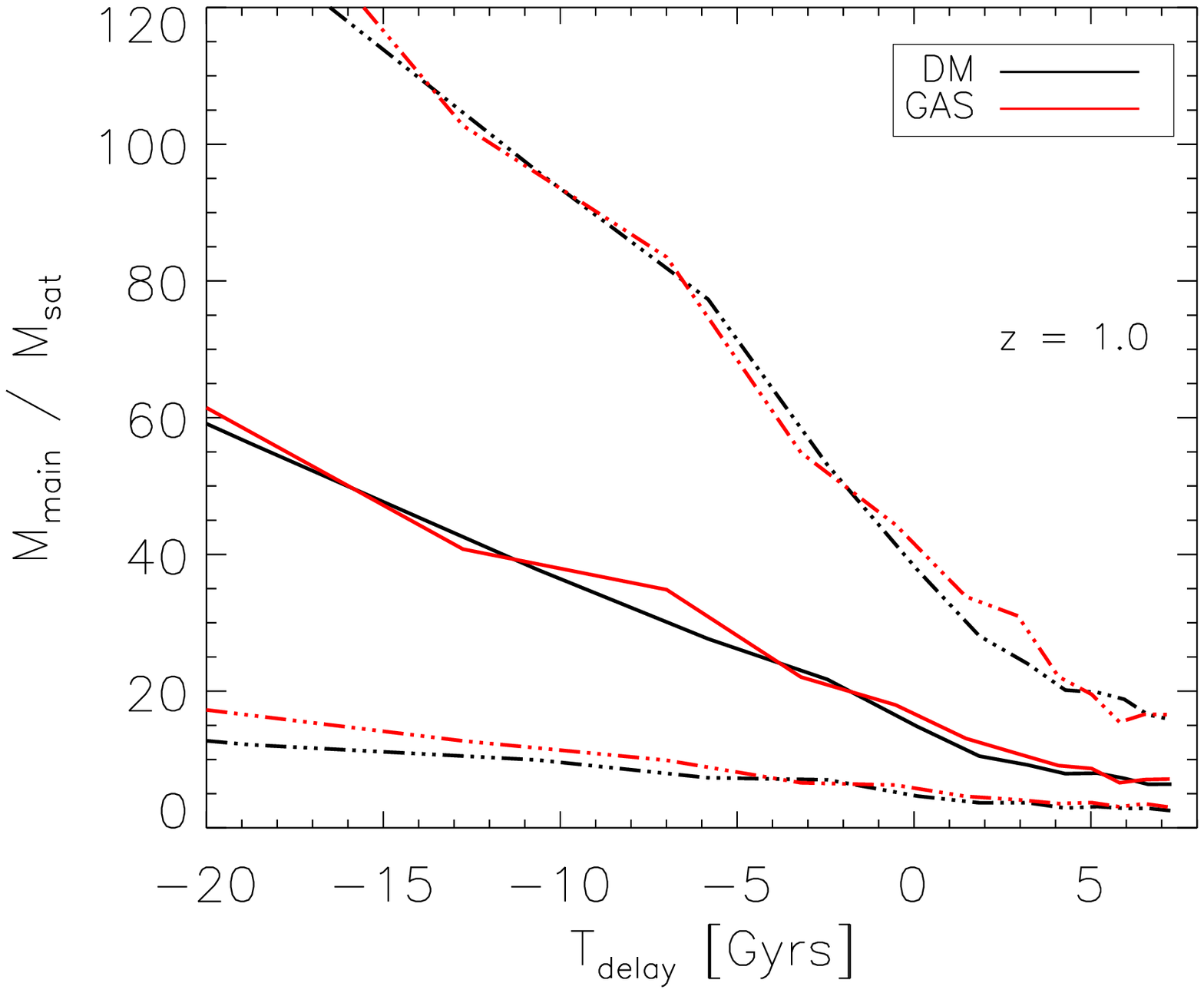,width=9.0cm} }}
  \caption{Dependence of satellite mass (upper left panel), distance from the
    centre of the accreting halo (upper right panel), circular velocity of the
    accreting halo at the virial radius (lower left panel), and ratio between
    the mass of the accreting halo and the satellite mass (lower right panel)
    on $T_{\rm delay}$. In all panels, solid lines show the median of the
    distributions, while dot-dashed lines show the 25th and 75th percentile of
    the distributions. Black lines are used for the DM runs and red lines for
    the GAS runs. Quantities refer to Type-2 galaxies identified at $z=1$.}
  \label{fi:merg_el} 
\end{figure*}

The upper left panel of Figure ~\ref{fi:merg_el} shows the dependence of the
satellite mass on $T_{\rm delay}$, for all Type-2 galaxies identified at $z=1$.
The solid lines show the median of the distributions, while dot-dashed lines
mark the 25th and 75th percentiles. Black and red lines are used for the DM and
the GAS runs respectively. The figure indicates that, for a fixed value of
$T_{\rm delay}$, Type-2 galaxies in the GAS runs are created in substructures
whose mass is systematically lower than the corresponding quantity in the DM
runs. This difference is of the order of the baryonic fraction, and is due to
the fact that when subhaloes lose their identity in the GAS runs, their
baryonic component has been stripped by ram--pressure. A systematically lower
mass for substructures in the GAS runs provides indeed longer merging times for
the Type-2 galaxies (see Eq.~\ref{eq:merg}).
 
No significant difference between the GAS and the DM runs can be noticed for
quantities like the distance between the merging halo and the centre of the
accreting halo (top right panel of Figure \ref{fi:merg_el}) and the circular
velocity of the accreting halo (bottom left panel of Figure \ref{fi:merg_el}).
Finally, the bottom right panel of Figure \ref{fi:merg_el} shows the ratio
between the mass of the accreting halo and the satellite mass, which enters in
the Coulomb logarithm at the denominator of Eq.~\ref{eq:merg}. If the effect of
ram--pressure was that of stripping gas only from the satellite and not from
the accreting halo, we would expect a difference between the GAS and the DM
runs, in the opposite sense of that shown in the top left panel of Figure
\ref{fi:merg_el}. The figure shows a very slight tendency for larger mass
ratios in the GAS runs. This suggests that also the accreting haloes tend to
have a deficit of gas, with respect to the DM runs, although the difference is
less significant than that found for the satellite mass. In addition, the
merging times have a logarithmic dependence on the mass ratio. 

\begin{figure}
  \centerline{ \hbox{ \psfig{file=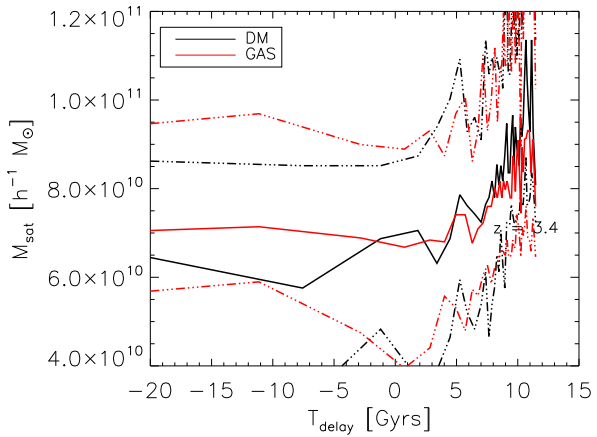,width=8.0cm} }}
  \caption{Satellite mass as a function of $T_{\rm delay}$ (as in the upper
    left panel of Figure \ref{fi:merg_el}), for all Type-2 galaxies identified
    at $z = 3.4$.}
  \label{fi:merg_z3} 
\end{figure}

Results shown in Figure \ref{fi:merg_el} show then that longer merging times
for Type-2 galaxies in the GAS runs are essentially due to a systematic
decrease of the satellite mass, caused by ram--pressure stripping. Figure
\ref{fi:merg_z3} shows again the satellite mass as a function of $T_{\rm
  delay}$, but this time for all Type-2 galaxies identified at $z \simeq
3.4$. At this earlier epoch, the cluster is not fully assembled yet, and the
proto-cluster region contains gas with lower pressure. It is to be expected
then that the effect of ram--pressure stripping is less significant. This
expectation is confirmed by the results shown in the figure, which suggests no
significant difference between the GAS and DM runs at this epoch. 
 
It is worth noticing at this point that the effect of ram--pressure stripping
is likely to be over-estimated in non--radiative simulations. In a more
realistic case, we expect a significant fraction of baryons to be converted
into stars before the effect of ram-pressure stripping becomes significant.
This expectation, however, needs to be verified with hydrodynamical simulations
which also include star formation and feedback processes.

%%%%%%%%%%%%%%%%%%%%%%%%%%%%%%%%%%%%%%%%%%%%%%%%%%%%%%%%%%%%%%%%%%%%%%%%%%%%%%%
\subsection{The brightest cluster galaxies}
\label{sec:BCGs}

In the previous sections, we have carried out a statistical comparison between
the galaxy populations from the DM and GAS runs. The general level of agreement
is quite good, although our analysis points out a number of interesting
differences. In particular, we have shown that the presence of the gas affects
the dynamics of subhaloes, so as to make an object-by-object comparison
difficult. This difficulty does not hold for the brightest cluster galaxies
(BCGs) which, due to their special location at the centre of the biggest halo
in the simulation, can be unambiguously identified both in the DM and in the
GAS runs.  In this section, we will compare the BCG formation history in the
two sets of runs used for our study.

\begin{figure*}
  \centerline{\hbox{\psfig{file=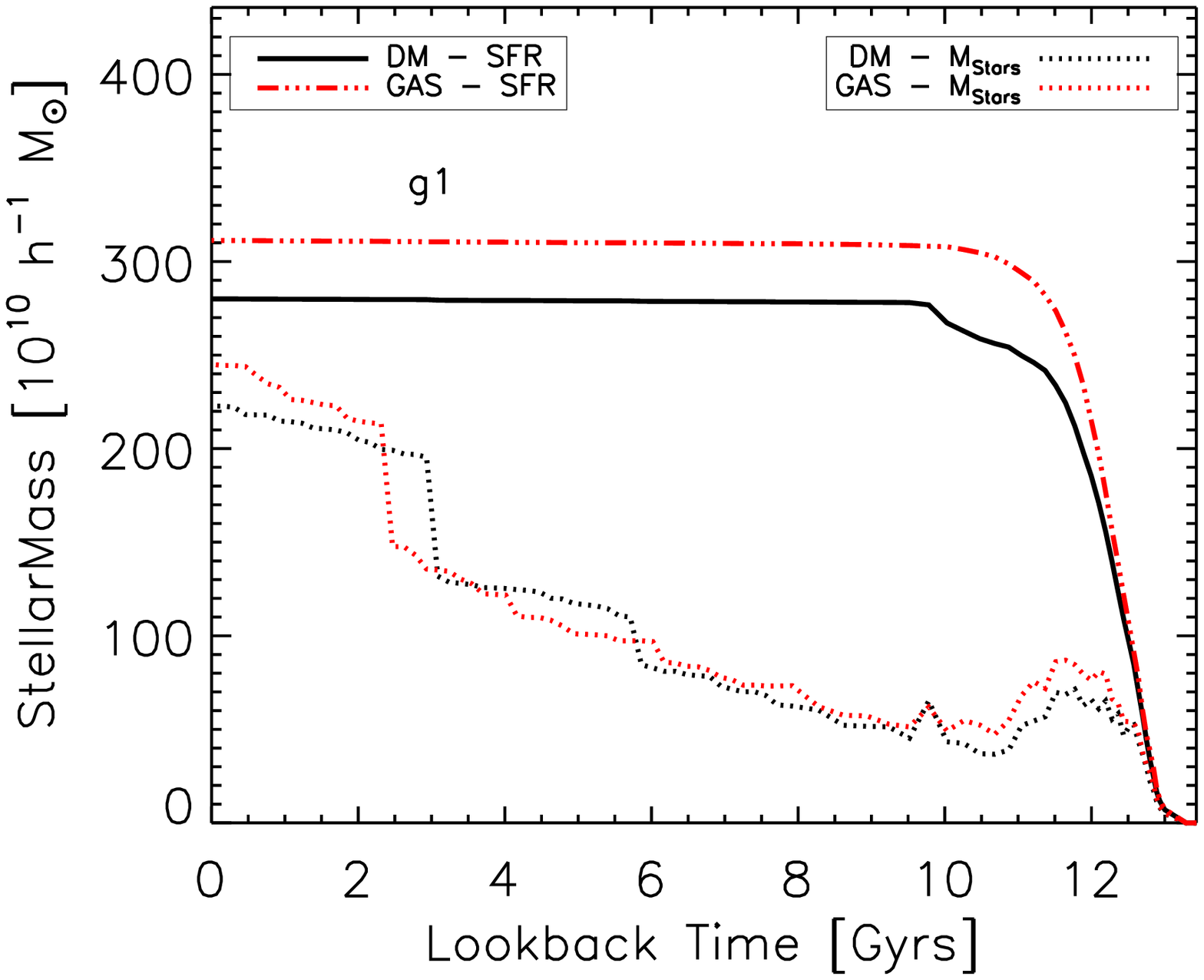,width=8.0cm}
      \psfig{file=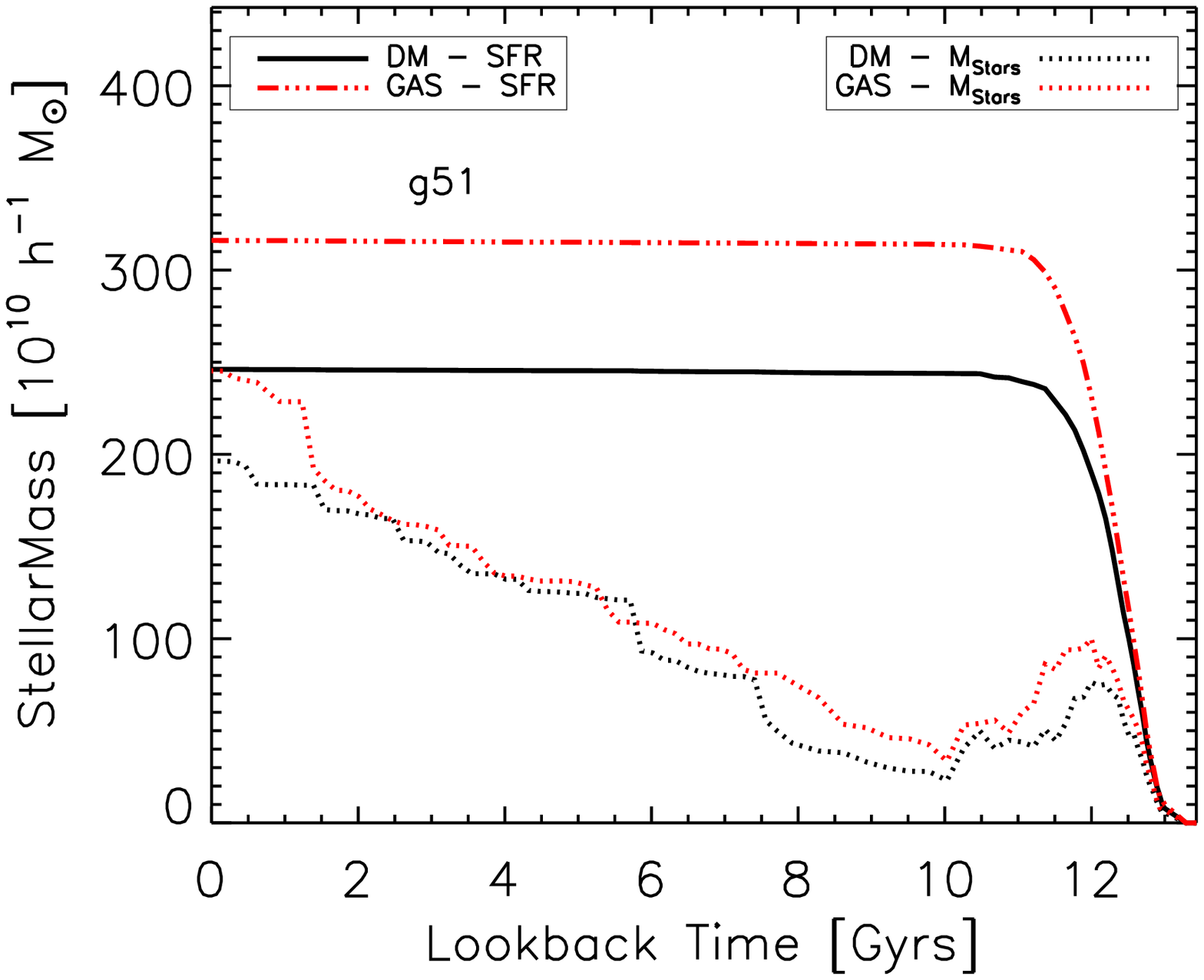,width=8.0cm} }} 
  \centerline{\hbox{\psfig{file=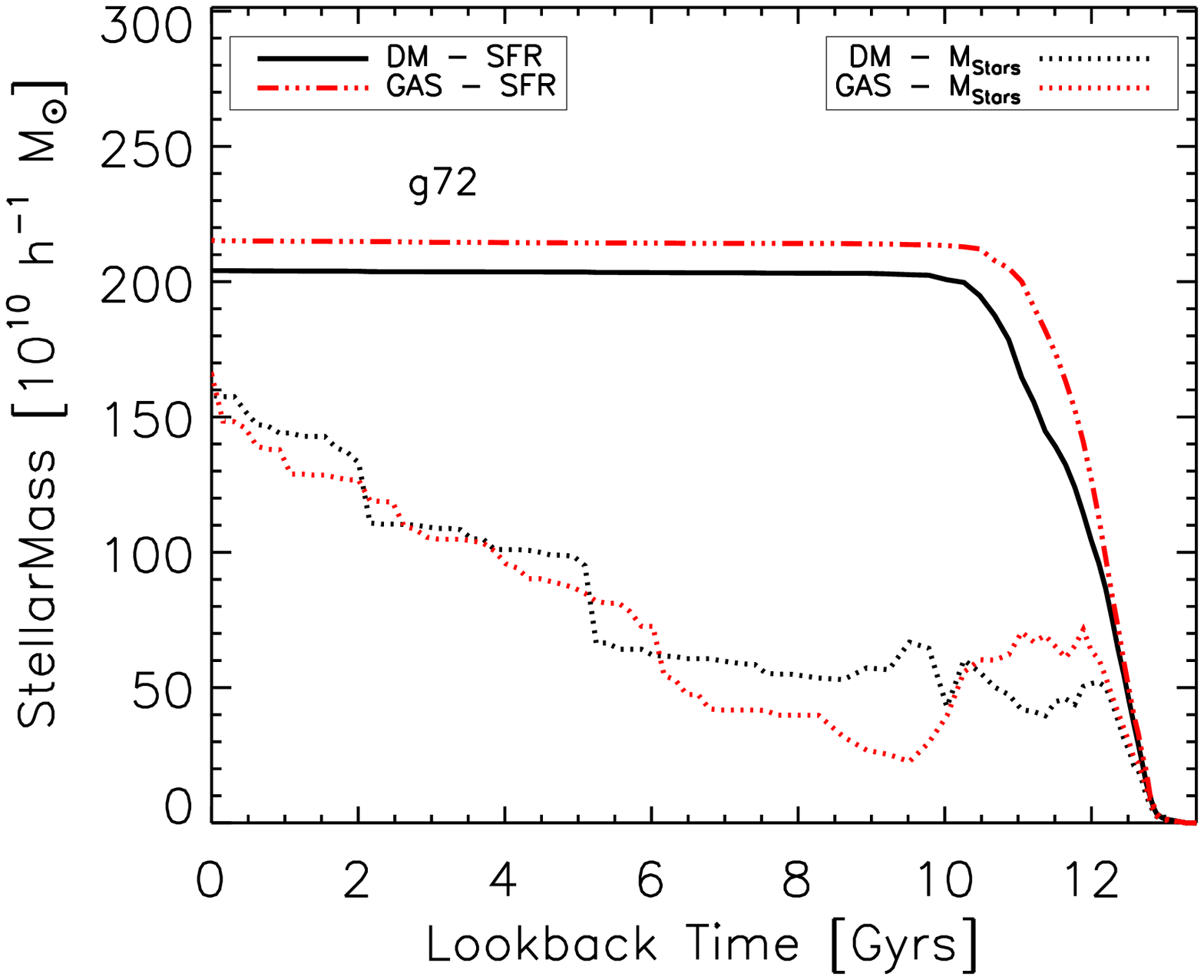,width=8.0cm}
      \psfig{file=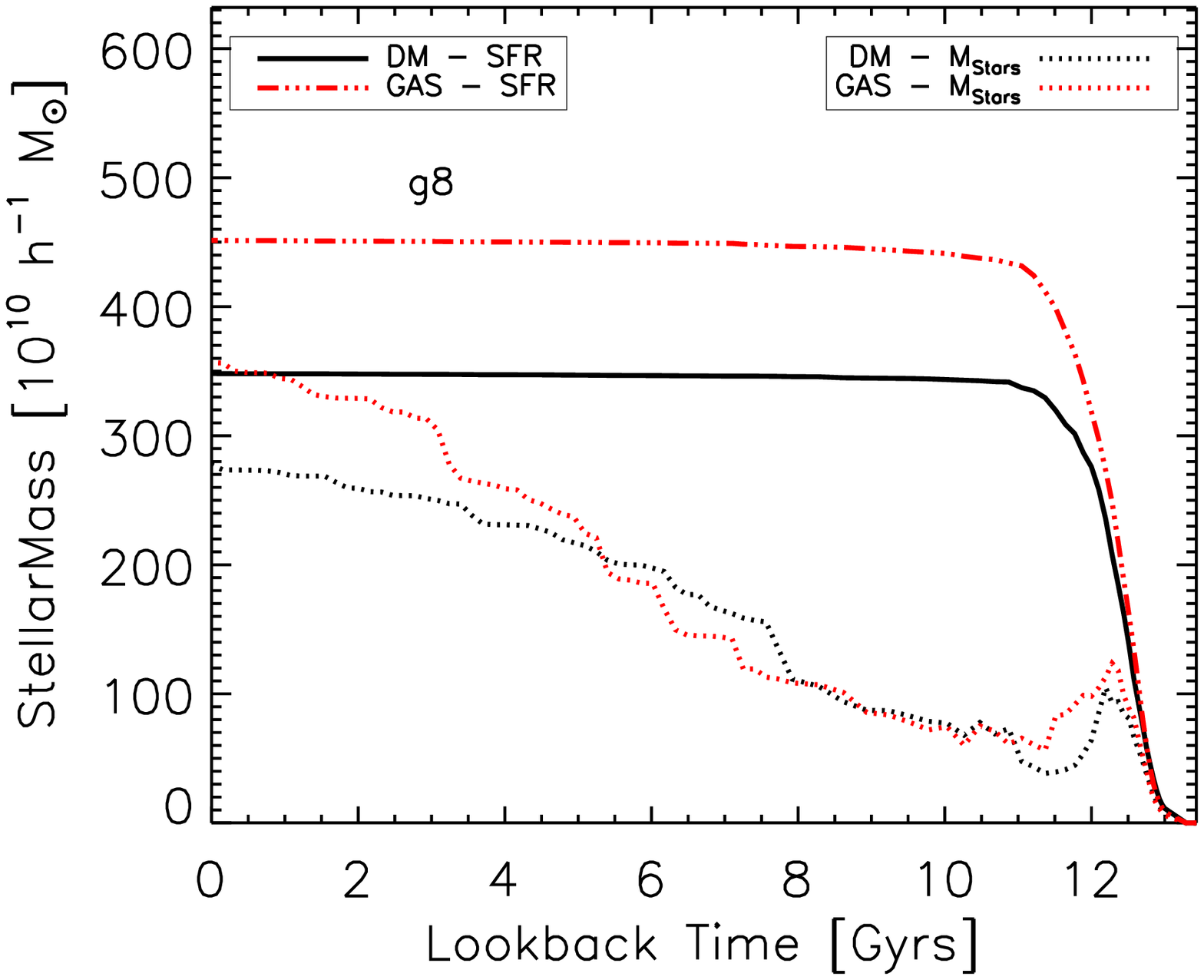,width=8.0cm} } }
  \caption{Evolution of the Type-0 progenitors of the BCG for the four
    simulated clusters used in this study. In each panel, the solid
    black lines and the dashed red lines show the integral of the SFR
    associated with the progenitors of the BCG for the DM runs and for
    the GAS runs respectively. The black and the red dotted lines show
    the total stellar mass of all Type-0 progenitors of the BCGs.}
  \label{fi:BCGs} 
\end{figure*}

For each of the simulated clusters used in our study, we constructed the full
merger tree of the final BCG tracing back in time all its progenitors and their
histories. In Figure \ref{fi:BCGs}, we show as dotted lines the total stellar
mass contained in the Type-0 progenitors of the BCG. This mass traces, for most
of the time, the stellar mass of the main progenitor of the BCG. At very early
times, it also includes the stellar mass of other central galaxies that belong
to the tree of the BCG and that are accreted onto the cluster halo at later
times (see Figure 1 of \citealt{2007MNRAS.375....2D}). We also show the
integral of the star formation rate (SFR) in all the Type-0 progenitors for the
DM runs (solid black lines) and for the GAS runs (dot-dashed red lines). The
difference between the final stellar mass and the integral of the SFR is due to
stellar mass losses\footnote{We recall that we adopt an instantaneous recycling
  approximation and that, for the adopted IMF, the recycled fraction is 0.3}.
In all the simulations used in this study, the integral of the SFR is constant
over the last 10 Gyrs, suggesting that all stars that end up in the final BCGs
where already formed at $z\simeq 2$, in agreement with findings by
\citet{2007MNRAS.375....2D}. Figure \ref{fi:BCGs} also shows that the BCGs in
the GAS runs are systematically more massive than their counter-parts in the DM
runs, confirming the visual impression from Figure \ref{fi:Maps}. This
difference amounts to a few per cent in the g72 simulation (lower left panel),
but it reaches values larger than $25$ per cent of the final stellar mass for
the g51 simulation (upper right panel). The difference is established at high
redshift, during the formation of the bulk of the stars that end up in the
final BCGs (i.e. the rising part of the curves showing the integral of the
SFR).

We have verified that this difference is sometimes due to one massive satellite
galaxy which merges with the main progenitor of the BCG within $z=0$ in the GAS
run, while it gets a longer time-scale for merging in the DM run. This can be
clearly seen in the case of the g51 cluster (upper right panel of Figure
\ref{fi:BCGs}), where the red dotted line shows an increase of about $5\times
10^{11}h^{-1}M_\odot$ in stellar mass due to a single merging event at a
lookback time of $\sim 1.5$~Gyr. Note that this behaviour is opposite to the
statistical trend that we have observed in section \ref{sec:Merg_time}, where
Type-2 galaxies in the GAS runs were found to have longer merging times.

\begin{figure*}
  \centerline{\hbox{ \psfig{file=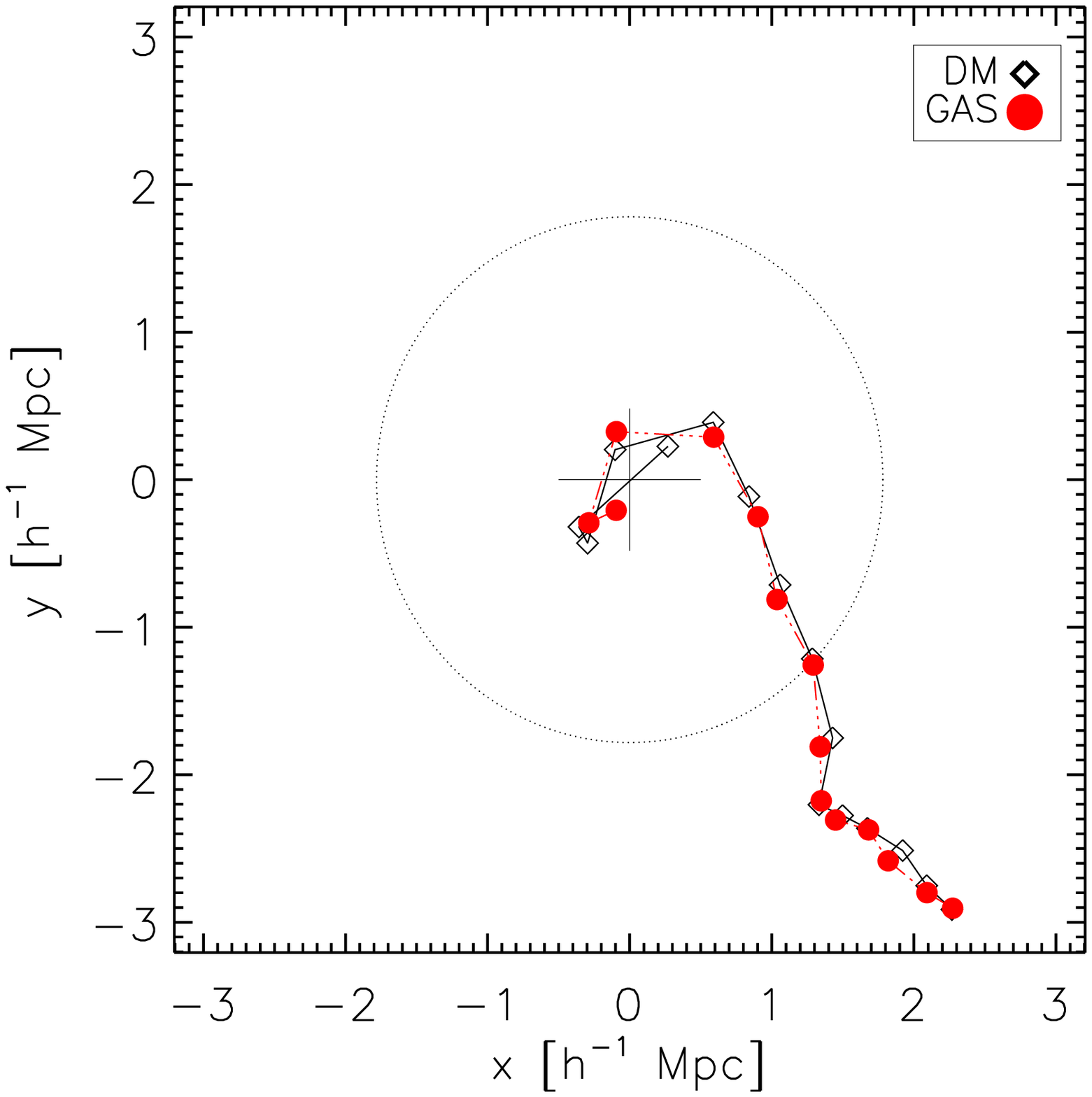,width=8.0cm}
      \psfig{file=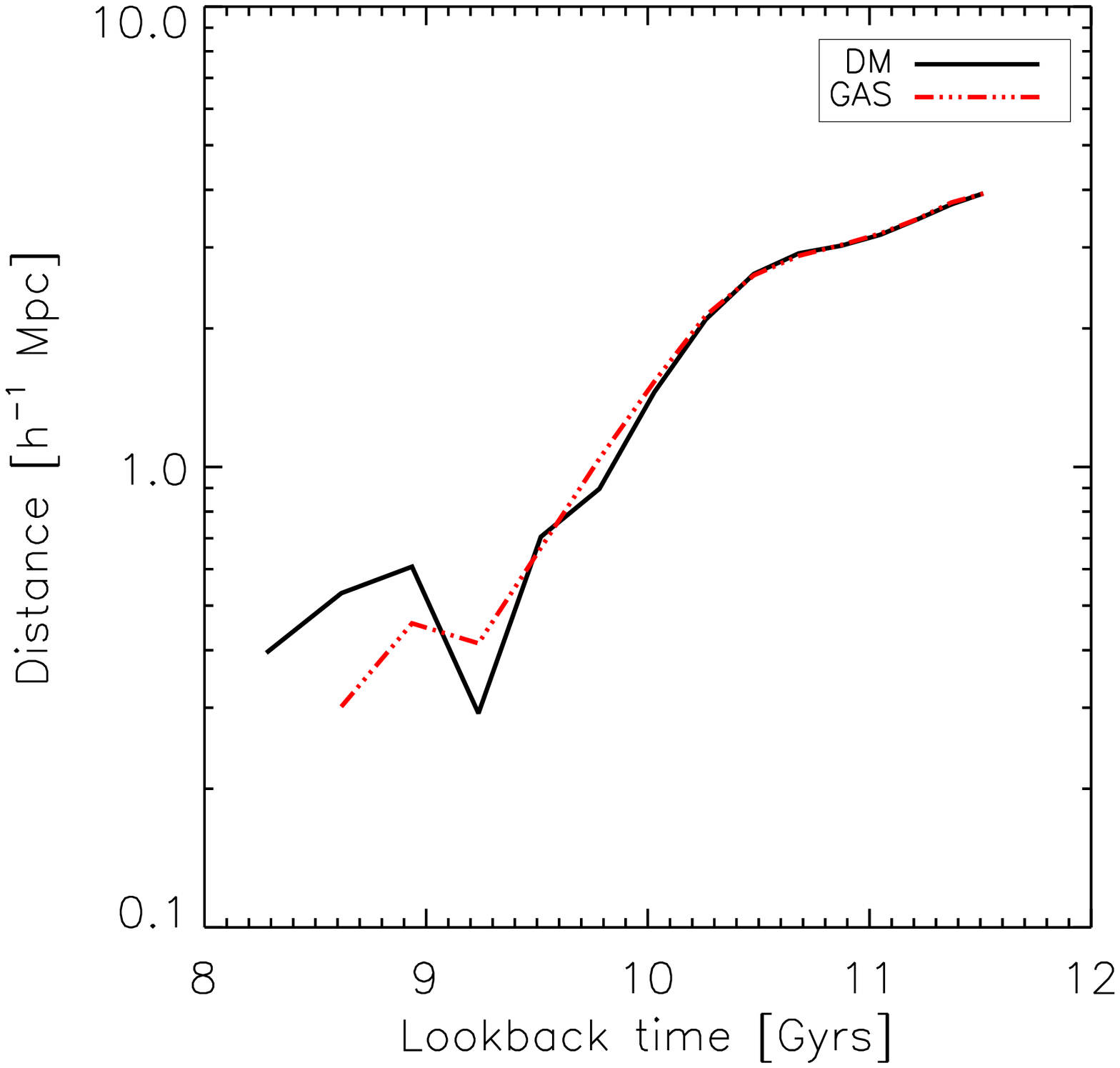,width=8.0cm} }}
  \caption{Left panel: comparison between the trajectories of a subhalo merging
    with the g51 cluster, projected on the $x$--$y$ plane. Open black diamonds
    are for the DM run, while filled red circles are for the GAS run. At each
    redshift, the coordinates of the merging subhalo are computed with respect
    to the cluster centre at the corresponding redshift. The big black
      circle shows the centre of the cluster at each snapshot, and its radius is equal
      to the corresponding value of $r_{200}$ at the last snapshot in
      which the subhalo was identified in the DM run. Right panel: evolution of the
    cluster--centric distance of the same subhalo in the DM (black solid line)
    and GAS (red dot-dashed line) run.}.
  \label{fi:traj} 
\end{figure*}

In order to understand in detail this difference, we show in the left
panel of Figure \ref{fi:traj} the orbit of the subhalo originating
this Type-2 galaxy, before its merging with the g51 cluster. Black
diamonds show the orbit from the the DM run, while red filled circles
correspond to the orbit from the GAS run.  The positions of the
subhalo are initially very similar in the two runs.  As it approaches
the high-density environment of the cluster, it is slowed down by
ram--pressure. The right panel of Figure \ref{fi:traj} shows the evolution of
the cluster-centric distance of the subhalo. The figure shows that the
pericentric and apocentric passages in the GAS runs take place at a
larger distance in the GAS run with respect to the DM run at a
lookback time of about 9.3 Gyr, and at a smaller distance at a
lookback time of about 8.9 Gyr. The subhalo has then a more circular
orbit in the GAS run \citep[see also][]{2005A&A...442..405P}. Besides
modifying the shape of the orbit and the timing of the merging,
ram--pressure also makes the substructure more fragile. Indeed, this
subhalo loses its identity 8.6 Gyr ago in the GAS run, at a
cluster-centric distance of about $0.3\hm$. In the DM run, the same
subhalo loses its identity about 8.3 Gyr ago, at a cluster-centric
distance of about $0.4\hm$. Since the residual merging time assigned
to the galaxy at its centre is proportional to the square of this
distance (see Eq.~\ref{eq:merg}), the resulting merging time in the
GAS run is more than 40 per cent shorter than in the DM run, and the
subhalo disappears about 0.3 Gyr earlier. The merging occurs before
$z=0$ in the GAS run, and causes the sudden increase of the stellar
mass of the BCG visible in the top right panel of Figure
\ref{fi:BCGs}.

\begin{figure}
 \centerline{\hbox{\psfig{file=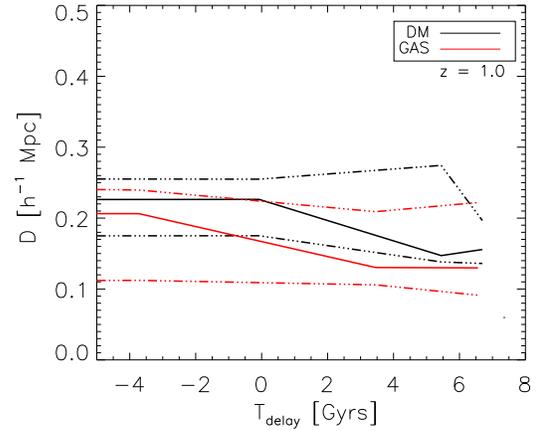,width=8.0cm}}} 
 \caption{Same as for the upper right panel of Figure \ref{fi:merg_el}, but
   only for Type-2 galaxies with stellar masses larger than $10^{11}\msun$. }
 \label{fi:dist11} 
\end{figure}

We recall that the upper right panel of Figure \ref{fi:merg_el} shows that
there is no significant difference, in terms of distance from the accreting
halo, between the DM and GAS runs. This is in apparent contradiction with the
above example. That figure was, however, obtained for all the Type-2 galaxies
identified at $z=1$, irrespective of their mass. In Figure \ref{fi:dist11}, we
repeat the same plot but considering only Type-2 galaxies with stellar masses
larger than $10^{11}\msun$. These massive satellites belong to subhaloes that
lose their identity at systematically smaller distances in the GAS runs, like
in the example illustrated in Figure \ref{fi:traj}. This example is then not
just a statistical fluctuation, but rather the result of a more general trend
for massive Type-2 satellites whose number is, however, quite low.

\begin{figure}
  \centerline{\hbox{
      \psfig{file=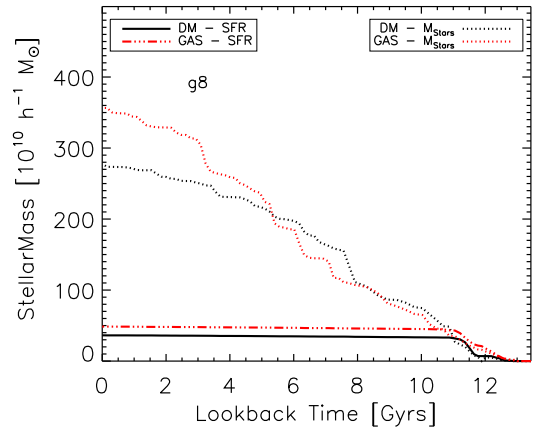,width=8.0cm} } }
  \caption{Total mass in the main progenitor of the BCG of the cluster g8
    (dotted lines) for the DM (black) and GAS (red) runs. The black solid line
    and the red dot-dashed line show the integral of the SFR in the main
    progenitor of the BCG.}
  \label{fi:first_prog} 
\end{figure}

Finally, we note that the difference between the mass of the BCGs in the DM and
GAS runs is due mainly to a different number of progenitors, rather than to a
difference in their intrinsic star formation rate. This is illustrated in
Figure \ref{fi:first_prog} which shows the amount of stars formed `in situ' in
the main progenitor of the BCG (solid black and dot-dashed red lines), and the
total stellar mass in the main progenitor at each time (dotted lines). The
stars formed in the main progenitor make up only a small fraction (about one
tenth) of the final stellar mass in the BCG, and most of these stars are formed
relatively early (more than 10 Gyrs ago). This is in agreement with results by
\citet[][see their Figure 4]{2007MNRAS.375....2D}. The figure shows that the
amount of stars formed `in situ' is comparable in the DM and GAS runs while the
total mass in the main progenitor of the BCG in the GAS run increases more
steeply than in the DM run, and reaches a final value that is about 1.3 times
larger.

%%%%%%%%%%%%%%%%%%%%%%%%%%%%%%%%%%%%%%%%%%%%%%%%%%%%%%%%%%%%%%%%%%%%%%%%%%%%%%%
\section{Conclusions and discussion}
\label{sec:Concl}

In this paper we have used numerical simulations to analyse how the
presence of non--radiative gas dynamics affects the predictions of
semi-analytic models of galaxy formation for the properties of cluster
galaxies. The main results of our work can be summarised as follows.

\begin{enumerate}
  
\item The stellar mass function of galaxies from DM-only runs is in quite good
  agreement with that obtained from non--radiative hydrodynamical runs. This
  result is a combination of two different and opposite effects.

  \begin{itemize}
  \item Due to a reduced number of subhaloes in the GAS runs (see Figure
    \ref{fi:MF_haloes}), these simulations result in a galaxy population with a
    reduced number of Type-0 and Type-1 galaxies (i.e. central galaxies of a
    halo, either the main halo or a proper substructure).
  \item Due to a systematic increase of the residual merging times assigned to
    Type-2 galaxies (those associated with haloes disrupted below the resolution
    limit of the simulation), the cluster galaxy population in the GAS runs
    contains a larger number of Type-2 galaxies than the DM runs. 
  \end{itemize}
  
\item The longer merging times assigned 
%SB.
on average 
to Type-2 galaxies
  in the GAS runs are due to ram--pressure stripping, which removes
  gas from the merging subhaloes and makes them more fragile. The
  effect of ram--pressure is more important at lower redshift, when
  the cluster has already assembled in a dominant structure with a
  high--pressure atmosphere that can efficiently remove gas from
  substructures. When considering the entire satellite population, we
  find a systematic difference between the DM and GAS runs in the
  sense that merging substructures are less massive in the runs with
  gas. This trend, however, is reversed when concentrating on the most
  massive satellites
%SB.
(see item iv below).
\vspace{0.1cm}
  
\item Type-2 galaxies dominate the radial density profile of cluster galaxies
  particularly in the inner regions, in agreement with results by
  \citet{Gao_etal_2004}. Galaxies associated with distinct dark matter
  substructures (Type-1 galaxies) exhibit a flatter distribution and their
  contribution to the inner regions of galaxy clusters is negligible. We did
  not find any significant difference, in terms of spatial distribution,
  between the DM and the GAS runs.\vspace{0.1cm} 

\item Although a statistical comparison between galaxy populations from the two
  sets of runs results in a quite nice agreement, a one-to-one comparison for
  the brightest central galaxies shows that these galaxies tend to have larger
  stellar masses in runs with gas. The difference varies from cluster to
  cluster and it is generally due to single merging events of relatively massive
  satellites which get assigned lower merging times in the GAS runs (see the
  example shown in Figure \ref{fi:traj}). The final difference in stellar mass
  is then due primarily to a different accretion history of satellite galaxies
  in the two sets of runs, and not to intrinsic differences in the star
  formation rates in the main progenitor.
\end{enumerate}

Our results demonstrate that predictions of semi-analytic models of
galaxy formation are not significantly affected when non-radiative
hydrodynamic simulations are used to construct the halo merger trees
which provide the skeleton of the model. This statement is, however,
correct only in a statistical sense. The presence of the gas induces
significant differences in the timing of the halo mergers, and affects
significantly the halo orbits making them more circular, on
average. Although these effects might be over-estimated in our
non--radiative runs, our results suggest that an accurate treatment of
merging times is crucial for predicted quantities like the mass
accretion history of model brightest cluster galaxies. As subhaloes
are fragile systems that are rapidly reduced below the resolution
limit of the simulation
\citep{2004MNRAS.348..333D,2004MNRAS.355..819G}, the treatment of
satellite mergers in semi-analytic models requires the use of analytic
formulations (e.g.  the Chandrasekhar formula). Recent work
\citep{2008MNRAS.383...93B,2008ApJ...675.1095J} has shown the limits
of the formulation usually adopted in semi-analytic models. This
recent work, however, does not provide consistent alternative
formulations. Additional work is therefore needed in order to obtain a
more realistic and detailed description of the merging process, which
represents a crucial ingredient of semi-analytic models of galaxy
formation.

\section*{Acknowledgements}
%SB. 
We thank Volker Springel for making available the substructure finder and
merger tree construction software that was originally developed for the
Millennium Simulation project. 
%SB.
We acknowledge useful discussions with Pierluigi Monaco and Giuseppe
Murante. 
AS acknowledges the receipt of a Marie Curie
Host Fellowships from the EARA-EST programme, and the hospitality of the
Max-Planck-Institut f\"ur Astrophysik where this project was initiated. GDL
acknowledges the receipt of a Short Visit Grant from the European Science
Foundation (ESF) for the activity entitled `Computational Astrophysics and
Cosmology', and the hospitality of the Dipartimento di Astronomia
dell'Universit\`a di Trieste where part of this work was carried out. 
%SB. 
We acknowledge financial contribution from the INFN-PD51 grant and the
contract ASI I/016/07/0 (COFIS).
\bsp

\label{lastpage}

\bibliographystyle{mn2e}
\bibliography{paper.bib}

\end{document}